\documentclass[journal]{IEEEtran}

\usepackage{graphicx}
\usepackage{subfigure}
\usepackage{balance}
\usepackage{colortbl}
\usepackage{multicol}
\usepackage{multirow}
\usepackage{tabularx}
\usepackage{mathdots}
\usepackage{colordvi}
\usepackage{color}
\usepackage{amsmath}
\usepackage{float}
\usepackage[none]{hyphenat}
\usepackage[noadjust]{cite}
\usepackage{soul}
\usepackage{amssymb}
\usepackage{pifont}
\newcommand{\cmark}{\ding{51}}%
\newcommand{\xmark}{\ding{55}}%

\definecolor{orange}{rgb}{1,0.5,0}

\newcommand{\fcro}[1]{\textcolor{black}{#1}}
\newcommand{\fcr}[1]{\textcolor{black}{#1}}

\newcommand{\diniz}[1]{\textcolor{black}{#1}}
\newcommand{\wam}[1]{\textcolor{black}{#1}}

\floatstyle{ruled}
\newfloat{algorithm}{!btp}{loa}
\providecommand{\algorithmname}{Algorithm}
\floatname{algorithm}{\protect\algorithmname}

\begin{document}
\title{Intersymbol and Intercarrier Interference in\\OFDM Systems: Unified Formulation and Analysis}

\author{Fernando~Cruz{--}Rold{\'a}n,~\IEEEmembership{Senior Member,~IEEE,}
         Wallace A. Martins,~\IEEEmembership{Senior Member,~IEEE},
		\\Fausto Garc\'ia G.,~
        Marc~Moonen,~\IEEEmembership{Fellow,~IEEE},
        Paulo S. R. Diniz,~\IEEEmembership{\diniz{Life} Fellow,~IEEE}
        

\thanks{F. Cruz-Rold{\'a}n and  F. Garc\'ia G. are with the Department of Teor{\'i}a de la Se{\~n}al y Comunicaciones, Escuela Polit{\'e}cnica Superior de la Universidad de Alcal{\'a}, 28871 Alcal{\'a} de Henares (Madrid), Spain (e-mail: fernando.cruz@uah.es).

\wam{W.A. Martins is with ISAE-SUPAERO, Universit{\'e} de Toulouse, 31055 Toulouse, France (e-mail: wallace.martins@isae-supaero.fr).}

\wam{P.S.R. Diniz is with the Electrical Engineering Program (PEE/Coppe) and the Department of Electronics and Computer Engineering (DEL/Poli), Federal University of Rio de Janeiro (UFRJ), 21941-972, Rio de Janeiro/RJ, Brazil (e-mail: diniz@smt.ufrj.br).}

M. Moonen is with the Department of Electrical Engineering (ESAT-STADIUS), KU Leuven, 3001 Leuven, Belgium (e-mail: Marc.Moonen@esat.kuleuven.be).}
   }

\maketitle

\begin{abstract}
\fcr{A new unified formulation for seven different orthogonal frequency-division multiplexing (OFDM) systems is presented. The proposed formulation relies on six parameters and encompasses conventional OFDM systems, with windowing in the transmitter and/or in the receiver, and also with \fcro{a} cyclic prefix (CP) or both \fcro{a} CP and cyclic suffix (CS). A new equivalent channel matrix that is useful for calculating both the received signal and the \fcro{intersymbol and intercarrier} interference power is defined and characterized.  Unlike previous works, this new channel matrix formulates the channel convolution with no restrictions on the length of the channel impulse response. Moreover, it includes the overlap-and-add procedure performed in the transmitter of windowed OFDM systems. Furthermore, new theoretical expressions for the intersymbol and intercarrier interference and also for the signal-to-interference-plus-noise ratio are derived. 
}
\end{abstract}


\IEEEpeerreviewmaketitle



\section{Introduction}

\PARstart{I}{n multicarrier modulation (MCM)} systems,  frequency-selective communication channels are effectively partitioned into a set of flat-fading channels, whose effects can be equalized by using  one coefficient per subcarrier. MCM can be implemented in several ways, but the most popular is orthogonal frequency-division multiplexing (OFDM) \cite{Bin90, Lin11, Cio16, Din12}.


\fcr{OFDM is standardized in the downlink of 4G and 5G, where it offers simplicity and effectiveness against frequency-selective fading, relative insensitivity to timing offsets, compatibility with multiple-input multiple-output (MIMO) systems, and high efficiency for enhanced-mobile broadband (eMBB) services. In the context of 5G, alternative \fcro{systems} or waveforms  \diniz{have been extensively studied for new scenarios \cite{Zha16, Wav5G}, such as massive machine-type communications (mMTC) or vehicular to everything (V2X).} Among these alternative \fcro{schemes}, windowed OFDM (w-OFDM)} \fcro{does} not have a severe negative impact on \fcr{the implementation of these new services}, unlike other \fcro{systems} such as wavelet OFDM or filter-bank multicarrier (FBMC)~\cite{Ach11, Cru15, Pin18, Pin21}, which are not fully compatible with the existing OFDM-based solutions.

\subsection{\fcr{Background}}

w-OFDM is a variation of OFDM that includes pulse-shaping or time-domain windowing. In addition, the windowed parts overlap with each other so as to reduce the time-domain overhead resulting from the windowing, achieving the same spectral efficiency as conventional cyclic-prefix OFDM (CP-OFDM). Due to its smooth transitions in the time domain, w-OFDM  reduces side-lobe levels and achieves better spectral efficiency, a higher reduction of the out-of-band (OOB) emission and/or adjacent channel interference (ACI) rejection, compared to conventional CP-OFDM. For these reasons, w-OFDM has been  widely deployed in several wireless and wireline communication standards (e.g., see \cite{Doc31, IEEE10, IEEE13, Ghn18, Ghn18b}).

\begin{table}
\center{}
\caption{Characteristics of the Considered OFDM Systems} 
{\begin{tabular}{ | c |  c |  c | c| c | c | }
  \hline
   \multirow{ 2}{*}{\textbf{System}} & \textbf{Windowing}  &  \multirow{ 2}{*}{\textbf{CP}} &  \multirow{ 2}{*}{\textbf{CS}} & \multirow{ 2}{*}{\fcr{\textbf{Example}}} \\
   & \textbf{Side} & & & \\
  \hline
  \hline
   CP-OFDM  &  \xmark & \cmark & \xmark & \cite{Lin11} \\
  \hline
   wtx-OFDM  &  Tx & \cmark  & \cmark &  \cite{DAl12} \\
   \hline
    wrx-OFDM  &  Rx  & \cmark & \cmark &  \cite{Mul01} \\
   \hline
   WOLA-OFDM  &  Tx/Rx & \cmark & \cmark &  \cite{Zay16}\\
   \hline
   CPW-OFDM  &   Tx/Rx & \cmark & \cmark &  \cite{An18}\\
   \hline
   CPwtx-OFDM  & Tx & \cmark & \xmark &  \cite{Ach11} \\
   \hline
   CPwrx-OFDM  & Rx & \cmark & \xmark &  \cite{Lin07}\\
   \hline
\end{tabular}}{\label{system_characteristics}}
\end{table}


\begin{table*}
\center{}
\caption{\fcr{Design Parameters Values for Proper System Operation Using a $\mu-$length Cyclic Prefix}.}  
{\begin{tabular}{ |  m{1.9cm}<{\centering}  | m{1.9cm}<{\centering} | m{1.9cm}<{\centering} | m{1.4cm}<{\centering} | m{2.6cm}<{\centering} | m{2.2cm}<{\centering} |}
  \hline
    \multirow{2}{*}{\textbf{System}}  &  $\beta$  &  $\delta$  &  $\rho$ &   $\gamma$ &   $\kappa$  \\
      &  \textbf{Tx window tail} &  \textbf{Rx window tail} &  \textbf{ CS length} &  \textbf{Rx removed samples } &  \textbf{ Rx circular shift}    \\
  \hline
  \hline
   \textbf{CP-OFDM}   & $0$ & $0$ & $0$ & $\mu$ & $0$  \\
   \hline
    \textbf{wtx-OFDM}  & $\beta < \mu$ & $0$ & $\beta$  & $\mu$ &  $0$ \\
   \hline
    \textbf{wrx-OFDM}   & $0$ & $\frac{\delta}{2} \leq \mu$ & $\frac{\delta}{2} $  & $\mu-\frac{\delta}{2} $  & $0$ \\
   \hline
    \textbf{WOLA-OFDM}  & $\beta < \mu - \delta$ & $\delta \leq \mu-\beta$ & $\beta $  &  $\mu-\delta$   & $\frac{\delta}{2}$  \\
   \hline
    \textbf{CPW-OFDM}  &  $\beta < \mu- \frac{\delta}{2}$ & $\frac{\delta}{2} \leq \mu-\beta$  & $\beta + \frac{\delta}{2}$ & $\mu- \frac{\delta}{2}$ & $0$ \\
   \hline
    \textbf{CPwtx-OFDM}  & $\beta < \frac{\mu}{2}$  & $0$ & $0$ & $\mu- \beta$  & $\beta$  \\
   \hline
    \textbf{CPwrx-OFDM}  & $0$  & $\delta \leq \mu$ & $0$  &  $\mu-\delta$  &   $\frac{\delta}{2}$ \\
   \hline
\end{tabular}}{\label{parameter_system_def_jun23}}
\end{table*}

\fcr{Different w-OFDM systems have been proposed in the literature} \cite{Mul01, Red02, Wei04, Lin07, Lin11, Ach11, DAl12, Bal13, Nha14, Wav5G, Zha16, Zay16, An18, Die19, Pek20, Hus22, Gim23}.  They comprise time-domain windowing in the transmitter (Tx) (e.g. \cite{DAl12}), which helps to control undesired OOB spectral components, i.e., to reduce spectral leakage.  Some systems also include a time-domain windowing in the receiver (Rx) to increase the OOB rejection and to reduce the power of interfering signals \cite{Mul01, Red02, Lin11, Pek20}, which can increase the signal-to-interference-plus-noise ratio (SINR). On the other hand, a \fcr{cyclic prefix (CP)} is always inserted in each transmitted data vector, \fcr{and an additional guard interval or cyclic suffix (CS) can also be appended \cite{Mul01, Wav5G, Zay16, An18, Gim23}}. 

\begin{figure}
\centering{
\vspace{-0.5cm}
\subfigure[\fcro{wtx-OFDM and WOLA-OFDM}]{\includegraphics[width=2.8in]{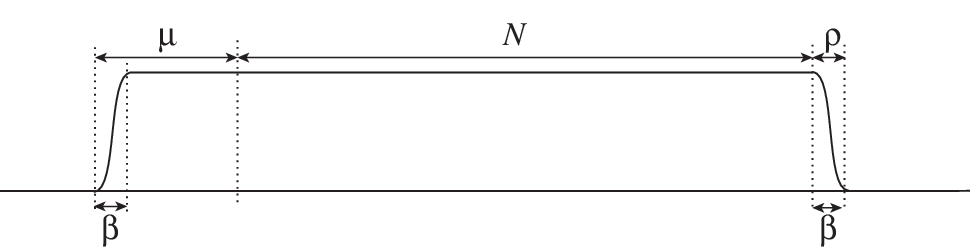} \label{Tx-Window_WOLA}}
\hfil 
\subfigure[\fcro{CPW-OFDM}]{\includegraphics[width=2.8in]{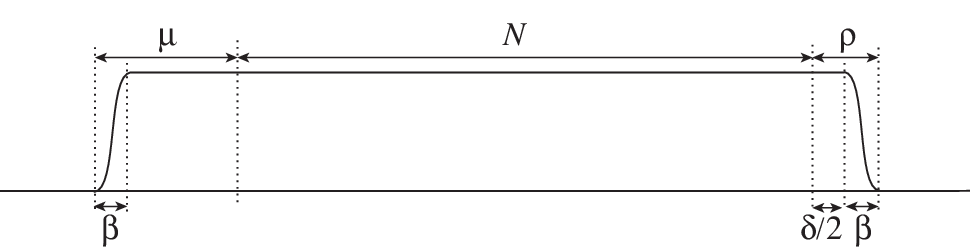} \label{Tx-Window_CPW}}
\hfil
\subfigure[\fcro{CPwtx-OFDM}]{\includegraphics[width=3.0in]{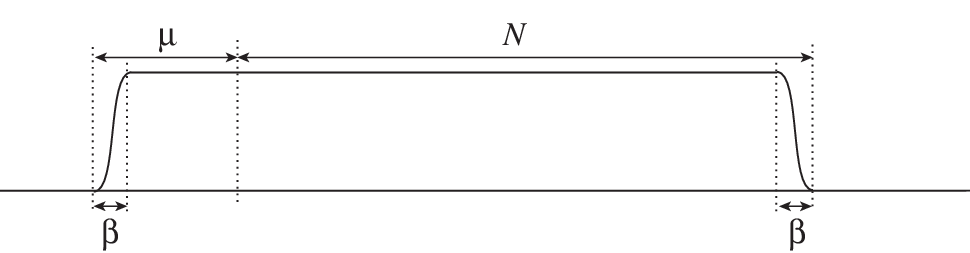} \label{Tx-Window_maderica}}
} \caption{\fcr{\wam{Tx} windows for different OFDM systems. $N$ is the number subcarriers, $\mu$ is the CP length, $\rho$ is the CS length, $\beta$ is the \wam{Tx} window tail, and $\delta$ is the Rx window tail.}} \label{fig:Tx-Window} 
\end{figure}
\begin{figure}
\centering{
\vspace{-0.5cm}
\subfigure[\fcro{wrx-OFDM and CPW-OFDM}]{\includegraphics[width=3.0in]{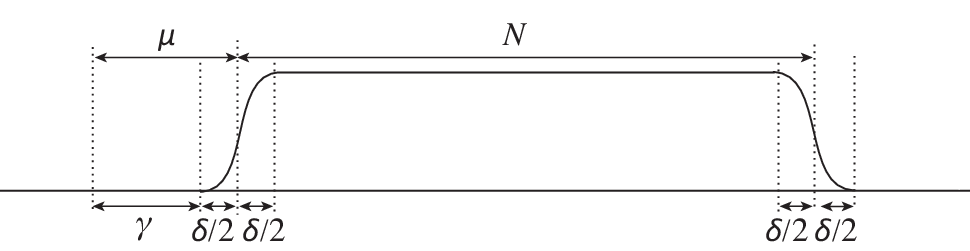} \label{Rx-Window_wtx}}
\hfil 
\subfigure[\fcro{WOLA-OFDM and CPwrx-OFDM}]{\includegraphics[width=2.8in]{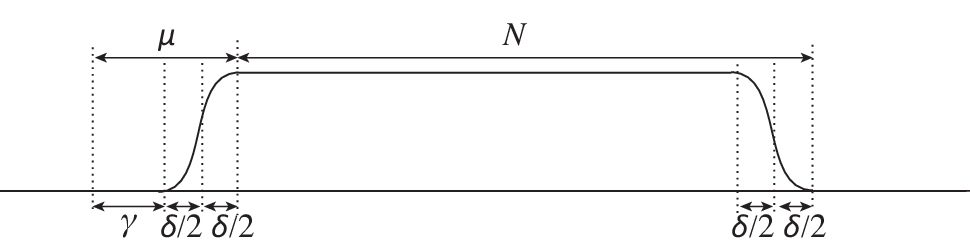} \label{Rx-Window_WOLA}}
} \caption{\fcr{\wam{Rx} windows for different OFDM systems. $N$ is the number subcarriers, $\mu$ is the CP length, $\gamma$ is the number of samples to be removed, and $\delta$ is the Rx window tail.} \label{fig:Rx-Window}} 
\end{figure}

\fcr{In this paper, we focus on seven different OFDM systems. First, we include in our study conventional CP-OFDM, which is the most widely standardized MCM \fcro{system}. Second, we consider OFDM with CP and CS, and time-domain windowing \fcro{either} only in the \fcro{Tx} (wtx-OFDM) or \fcro{only} in the Rx (wrx-OFDM) \cite{Wei04, Lin07, Lin11, Bal13, Die19, Gim23}}. Third, our study also includes the weighted overlap-and-add OFDM (WOLA-OFDM) \cite{Wav5G, Zha16, Zay16}, also with an additionally extended suffix \fcro{system} named CPW-OFDM \cite{An18}. \fcr{These} transceivers use independent time-domain windows in both \fcro{the Tx and Rx}, and \fcr{each transmitted data vector includes CP and CS}. Lastly, as there are standards that employ in their physical layers w-OFDM without CS, e.g., \cite{IEEE10, IEEE13, Ghn18}, it is necessary to study a fourth group of OFDM systems \fcr{that only \fcro{include a} CP, and windowing in the Tx (CPwtx-OFDM) or in the Rx (CPwrx-OFDM)~\cite{Ach11,Nha14}. Table \ref{system_characteristics} summarizes the characteristics of the considered OFDM systems, Table \ref{parameter_system_def_jun23} shows the design parameter values for proper system operation and, finally, Figs.~\ref{fig:Tx-Window} and ~\ref{fig:Rx-Window} depict the Tx and the Rx windows, respectively.} 

\subsection{\fcr{Interference in OFDM Systems}}
OFDM systems suffer from intersymbol and intercarrier interference (ISI and ICI) \fcr{when the CP length does not satisfy the conditions related to the order of the channel impulse response (CIR) \fcro{as} given in Table~\ref{no_interference}}. In this context, the interference analysis in conventional CP-OFDM has been widely addressed. For instance, different SINR models are derived in \cite{AlD96, AlD97, Kim98, Ars01, Ack01, Hen02, Mil02, Pha17, Lim17}. For more details, we refer the reader to \cite{Mar19}, where the impact of highly dispersive channels on OFDM under finite-duration CIR with arbitrary length is shown. Regarding w-OFDM, previous studies focusing on the analysis of interference are \cite{Bea07, Ach11, DAl12, Nha14, Pek20}. In \cite{Ach11, DAl12, Nha14}, both ISI and ICI are grouped under a single time-domain term, and the systems analyzed in these papers are CPwtx-OFDM \cite{Ach11, DAl12, Nha14} with a unique windowing in the \fcro{Tx}, and wtx-OFDM \cite{DAl12}.  In \cite{Pek20}, an Rx windowing OFDM system is considered, and the ICI induced by the proposed windowing is obtained.  \fcr{In our study, ICI and ISI  are comprehensively \wam{analyzed for} the seven different \fcro{systems} considered}. Since no constraint is imposed upon the order of the CIR, our results \fcr{are} applicable to the cases where the interference is due to any number of transmitted data blocks. 

\begin{table}
\centering
\caption{\fcr{Conditions to Avoid Interference Over a $\nu-$order Channel Impulse Response.}}
\begin{tabular}{|c|c|} \hline
    {\textbf{System}} & \textbf{Cyclic Prefix Length} \\ \hline \hline
    wtx-OFDM &  $\mu \geq \nu + \beta$ \\ \hline
    wrx-OFDM &  $\mu \geq \nu +\frac{\delta}{2}$ \\ \hline
    WOLA-OFDM &  $\mu \geq \nu + \beta + \delta$\\ \hline
    CPW-OFDM &  $\mu \geq \nu + \beta +\frac{\delta}{2}$\\ \hline
    CPwtx-OFDM & $\mu \geq \nu + 2\beta$\\ \hline
    CPwrx-OFDM & $\mu \geq \nu + \delta$\\ \hline
\end{tabular}
\label{no_interference}
\end{table}

\vspace{-0.5cm}
 \begin{figure}
 \centering{\includegraphics[scale=0.45]{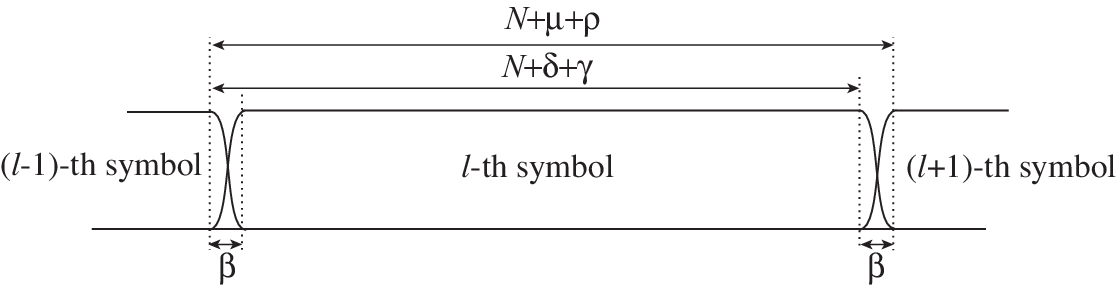}}
 \caption{\fcr{
Overlap-and-add at the transmitter side.}}
 \label{Overlap_and_add_tx}
 \end{figure}

\subsection{\fcr{Contributions}}
The main contributions in this paper can be summarized as follows:
\begin{itemize}
\item A \fcro{unified formulation} for a wide range of OFDM systems is presented. It includes the full transmission chain, the overlap-and-add operation \fcr{(see Fig. \ref{Overlap_and_add_tx})}, the convolution with a channel of arbitrary length, and the reception process. 
\fcr{This compact formulation based on a small set of only six parameters is practical\fcro{, for instance,} for \wam{software-defined} radio since it allows easy reconfiguration\fcro{, where} different w-OFDM systems are obtained by simply changing the parameter values}. In addition, it has excellent potential for use in an educational context, because it enables quickly explaining several OFDM systems \fcr{under unified expressions}. 
\item Theoretical closed-form expressions of \fcro{intersymbol and intercarrier} interference and noise for the case of an insufficient length of redundant samples are obtained. The interference is identified in the frequency domain, where the symbol is reconstructed, and classified into three different classes. This classification helps to study which \fcro{class} is most harmful to the system's performance.
\item Interference and noise powers  are derived to obtain the SINR, and hence the data rate and the \fcro{symbol-error rate (SER)}. \fcr{The resulting expressions are useful for window design,} bit-loading, adaptive CP and power/subcarrier allocation algorithms~\cite{Cio16}.
\end{itemize}

\subsection{\fcr{Organization and Notation}}

The rest of this paper is organized as follows. In Section~\ref{System_model}, we present the unified system model, considering seven different OFDM systems and adopting a \fcro{unified formulation} covering all of them. Then, three types of interference are calculated in Section~\ref{analyis_of_interfer}. In addition, theoretical expressions for both interference and noise powers are derived, and the corresponding SINR is determined. Simulation results are presented in Section~\ref{sec:experiments}, and finally, conclusions are drawn in Section~\ref{sec:conclusion}.

The notation used in this paper is as follows. Bold-face letters indicate vectors (lower case) and matrices (upper case). The transpose of ${\bf{A}}$ is denoted by ${\bf{A}}^T$ and ${\mathbf{I}}_N$ represents the $N \times N$ identity matrix. The subscript is omitted whenever the size is clear from the context.  
$\mathbf{0}$ and $\mathbf{1}$ denote, respectively, a matrix of zeros or ones.

\begin{figure*}
\centering{\includegraphics[width=0.8\linewidth]{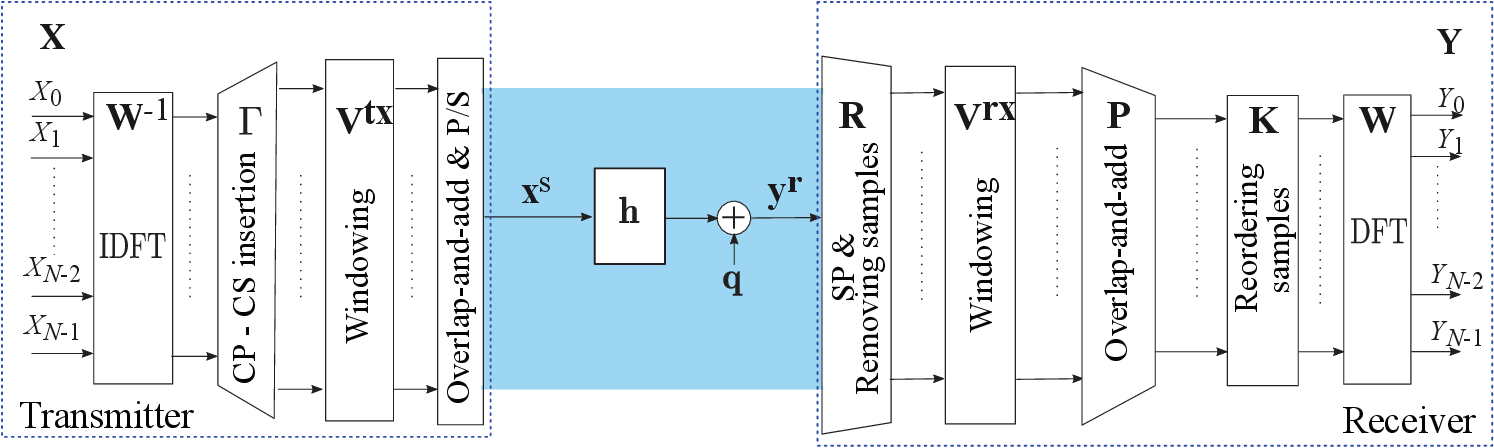}}
\caption{General block diagram of windowed OFDM over a channel with additive noise.}\label{block_diag}
\end{figure*}

\section{Unified \fcro{Formulation}} \label{System_model}
A block diagram is shown in Fig. \ref{block_diag}, where the transmitted data vector in the transform domain is given by
\begin{equation}
\mathbf{X}=\left[\begin{array}{cccc}
X_{0} & X_{1} & \cdots & X_{N-1}\end{array}\right]^{T},\label{eq:1}
\end{equation}
with $N$ being the number of subcarriers. The parameters used in the equations are \fcr{defined} in Table \ref{parameter_system_def_jun23}. We assume perfect synchronization in time and frequency, and also that the receiver has perfect channel-state information (CSI).

\subsection{Transmitter}
The \wam{$l$-th} time-domain \fcro{signal} vector before the overlap-and-add block is
\fcr{\[
\mathbf{x}_{(N+\mu+\rho) \times 1}^{\rm s} \left[ l \right] = \mathbf{V}_{(N+\mu+\rho)}^{\rm tx} \cdot \mathbf{\Gamma}_{(N+\mu+\rho) \times N} \cdot \mathbf{W}_{N}^{-1} \cdot \mathbf{X}_{N \times 1} \left[ l \right],
\]}
\noindent \fcro{where the matrices are defined as follows. First, $\mathbf{W}_{N}^{-1}$} represents the inverse DFT matrix with the $(k,n)$-th entry given by
\fcro{\[
\left[ {\bf{W}}_{N}^{-1} \right]_{k,n}  = \frac{1}{N} {\rm e}^{{\rm j}\frac{{2\pi }}{N}kn}, \quad 0 \leq k,n \leq N-1.
\]}
The matrix $\bf{\Gamma}$ introduces $\mu + \rho$ redundant samples:
\[
\bf{\Gamma}  = \left[ {\begin{array}{*{20}c}
   {\begin{array}{*{20}c}
   {{\bf{0}}_{\mu  \times (N - \mu )} } & {{\bf{I}}_{\mu  } }  \\
\end{array}}  \\
   {{\bf{I}}_{N} }  \\
   {\begin{array}{*{20}c}
   {{\bf{I}}_{\rho} } & {{\bf{0}}_{\rho  \times ({N} - \rho )} }  \\
\end{array}}  \\
\end{array}} \right].
\]
It appends a $\mu$-length CP and, when applicable, also a $\rho$-length CS. Observe that a cyclic shift, as employed in \cite{Ghn18b}, is equivalent to the inclusion of a CS into each data vector. \fcro{The windowing is represented by the} diagonal matrix 
\[
\mathbf{V}_{(N+\mu+\rho)}^{\rm tx}={\rm diag} \left\{\mathbf{v}_{1 \times (N+\mu +\rho)}^{\rm tx} \right\}, 
\]
obtained with a tapering window function, defined as
\[
\mathbf{v}_{1 \times (N+\mu+\rho)}^{\rm tx}=\left[\begin{array}{ccc}
\mathbf{v}_{1\times \beta}^{\rm tr} & \mathbf{1}_{1\times(N+\mu+\rho-2\beta)} & \mathbf{v}_{1\times \beta}^{\rm tf}\end{array}\right].
\]
The vectors $\mathbf{v}_{1\times \beta}^{\rm tr}$ and $\mathbf{v}_{1\times \beta}^{\rm tf}$ have as entries the rise and fall samples of the window tails, respectively.

After the pulse shaping or windowing, there is a $\beta$-samples overlap-and-add operation between successive symbols \fcr{, as is depicted in Fig. \ref{Overlap_and_add_tx}.} \fcro{In the next subsection, this} operation is jointly formulated with the channel convolution.

\subsection{Channel}
The signal $\mathbf{x}^{\rm s}$ is convolved with the transmission channel, defined as $
\mathbf{h}=\left[\begin{array}{cccc}
h_{0} & h_{1} & \cdots & h_{\nu}\end{array}\right],
$ and becomes contaminated by noise. 
In general, the number of transmitted data vectors that affect the first $N + \delta + \gamma$ samples of the received data vector is $M+1$, with
\begin{equation}
M \buildrel \Delta \over = \left\lceil{\frac{\nu + \beta}{N+\delta+ \gamma}}\right\rceil,
\label{eq:M-def}
\end{equation}
in which $\left\lceil \cdot \right\rceil $ represents the ceiling function. Therefore, the $l$-th received signal vector is given by
\begin{align*}
{\bf{y}}_{(N + \delta +\gamma) \times 1}^{\rm r} & \left[ {l} \right]  \\
 & = \sum\limits_{m =  0}^M {{\bf{H}}_{(N + \delta +\gamma) \times (N + \mu +\rho)}^{\left( { m} \right)} \cdot{\bf{x}}_{(N + \mu +\rho) \times 1}^{\rm s} \left[ {l -m} \right]} \nonumber \\ & + {\bf{q}}_{(N + \delta +\gamma) \times 1}[l] , 
\end{align*}
where ${\bf{H}}^{(m)}$ is a matrix whose entries, for $0 \leq b \leq N +\delta +\gamma -1$ and $0 \leq c \leq N +\mu +\rho -1$, are
\begin{equation}
\left[ {{\bf{H}}^{( m)} } \right]_{b,c}  \buildrel \Delta \over = \left\{ {\begin{array}{*{20}c}
   {0,} & {mN_0  + b - c < 0,}  \\
   {h_{mN_0  + b - c} ,} & {0 \le mN_0  + b - c  \le \nu ,}  \\
   {0,} & {mN_0  + b - c > \nu ,}  \\
\end{array}} \right.
\label{eq:H-def}
\end{equation}
where $N_0=N+\mu+\rho-\beta$ and $\mathbf{q}$ represents the channel noise.

\subsection{w-OFDM Receiver}
\fcr{The} received data vector can be expressed in the transform domain as
\begin{align}\label{received_signal_time_domain}
\mathbf{Y}_{N \times 1} \left[ l \right] &=\mathbf{W}_{N} \cdot \mathbf{K}_{N} \cdot \mathbf{P}_{N \times (N+\delta)} \cdot \mathbf{V}_{(N+\delta)}^{\rm rx} \nonumber \\
& \times   \mathbf{R}_{(N+\delta) \times (N+\delta+\gamma)} \cdot \mathbf{y}_{(N+\delta+\gamma) \times 1}^{\rm r} \left[ l \right], 
\end{align}
\fcro{where the} matrices are defined as follows. First, $\mathbf{R}$ represents removal of the first $\gamma$ samples of the received data vector:
\[
{\mathbf{R}} =\left[\begin{array}{ccc}
\mathbf{0}_{(N+\delta) \times \gamma} & \mathbf{I}_{(N+\delta)}  \end{array}\right].
\]

\noindent \fcro{The diagonal} matrix representing the windowing is
\[
\mathbf{V}^{\rm rx} = {\rm diag}\left\{\mathbf{v}_{1 \times (N+\delta)}^{\rm rx} \right\}, 
\]
where the tapering window in the Rx is defined as
\[
\mathbf{v}^{\rm rx} = \left[\begin{array}{ccc}
\mathbf{v}_{1\times \delta}^{\rm rr}  & \mathbf{1}_{1\times (N-\delta)} & \mathbf{v}_{1\times \delta}^{\rm rf}\end{array}\right],
\]
\noindent where $\mathbf{v}^{\rm rr}$ and $\mathbf{v}^{\rm rf}$ have as entries the rise and fall samples of the Rx window tails. Next, $\mathbf{P}$ is a matrix that represents a $\delta$-samples overlap-and-add operation: 
\[
{\bf{P}} = \left[ {\begin{array}{*{20}c}
   {\begin{array}{*{20}c}
   {{\bf{0}}_{{\delta  \mathord{\left/
 {\vphantom {\delta  2}} \right.
 \kern-\nulldelimiterspace} 2}} } & {{\bf{I}}_{{\delta  \mathord{\left/
 {\vphantom {\delta  2}} \right.
 \kern-\nulldelimiterspace} 2}} } & {{\bf{0}}_{{\delta  \mathord{\left/
 {\vphantom {\delta  2}} \right.
 \kern-\nulldelimiterspace} 2} \times \left( {N - \delta } \right)} } & {{\bf{0}}_{{\delta  \mathord{\left/
 {\vphantom {\delta  2}} \right.
 \kern-\nulldelimiterspace} 2}} } & {{\bf{I}}_{{\delta  \mathord{\left/
 {\vphantom {\delta  2}} \right.
 \kern-\nulldelimiterspace} 2}} }  \\
\end{array}}  \\
   {}  \\
   {\begin{array}{*{20}c}
   {{\bf{0}}_{\left( {N - \delta } \right) \times \delta } } & {{\bf{I}}_{N - \delta } } & {{\bf{0}}_{\left( {N - \delta } \right) \times \delta } }  \\
\end{array}}  \\
   {}  \\
   {\begin{array}{*{20}c}
   {{\bf{I}}_{{\delta  \mathord{\left/
 {\vphantom {\delta  2}} \right.
 \kern-\nulldelimiterspace} 2}} } & {{\bf{0}}_{{\delta  \mathord{\left/
 {\vphantom {\delta  2}} \right.
 \kern-\nulldelimiterspace} 2}} } & {{\bf{0}}_{{\delta  \mathord{\left/
 {\vphantom {\delta  2}} \right.
 \kern-\nulldelimiterspace} 2} \times \left( {N - \delta } \right)} } & {{\bf{I}}_{{\delta  \mathord{\left/
 {\vphantom {\delta  2}} \right.
 \kern-\nulldelimiterspace} 2}} } & {{\bf{0}}_{{\delta  \mathord{\left/
 {\vphantom {\delta  2}} \right.
 \kern-\nulldelimiterspace} 2}} }  \\
\end{array}}  \\
\end{array}} \right].
\]
Basically, it adds the first $\delta$ samples to the last $\delta$ samples. Then, a circular shift of $\kappa$ samples is needed in some systems (WOLA, CPwtx\wam{,} and CPwrx). This operation is formulated with the matrix $\mathbf{K}_{N}$, defined as follows:
\[
{\bf{K}}_N  = \left[ {\begin{array}{*{20}c}
   {{\bf{0}}_{(N - \kappa ) \times \kappa } } & {{\bf{I}}_{N - \kappa } }  \\
   {{\bf{I}}_\kappa  } & {{\bf{0}}_{\kappa  \times (N - \kappa )} }  \\
\end{array}} \right].
\]
In some other systems, e.g., those that include a CS in each transmitted data vector (wtx and CPW), this is an identity matrix: $\mathbf{K}_{N}=\mathbf{I}_{N}$. 
Finally, \fcro{${\bf{W}}_{N}$} is a DFT matrix:
\fcro{\[
\left[ {\bf{W}}_{N} \right]_{k,n}  = {\rm e}^{-{\rm j}\frac{{2\pi }}{N}kn}, \quad 0 \leq k,n \leq N-1. 
\]}

\section{Analysis of Interference}\label{analyis_of_interfer}
\begin{figure}\centering
\includegraphics[scale=1]{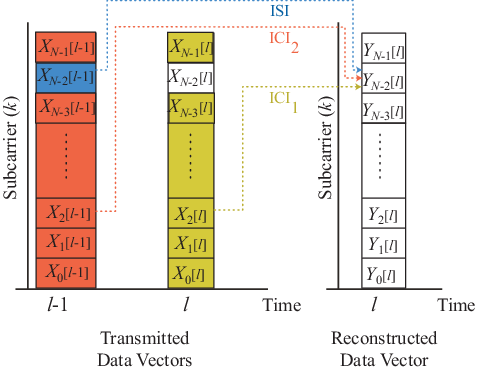}
\caption{Types of interference.}\label{fig:Tipos-de-interferencias}
\end{figure}

This section provides a comprehensive analysis of the interference and its power for the considered OFDM systems. Given a finite duration CIR with arbitrary length, without any constraint on the length, the received data vector can be expressed in the transform domain as
\begin{equation}\label{Eq_general_Yn}
{\bf{Y}}[l]  = \sum\limits_{m =  0}^M {{\bf{A}}_m  \cdot {\bf{X}}[l - m]} + {\bf{G}}^{\rm noise}  \cdot {\bf{q}}[l],
\end{equation}
where
\[
{\bf{A}}_{m, N \times N} =\mathbf{W} \cdot \mathbf{K} \cdot \mathbf{P} \cdot \mathbf{V}^{\rm rx} \cdot \mathbf{R} \cdot {\bf{H}}^{\left( m \right)} \cdot \mathbf{V}^{\rm tx} \cdot \mathbf{\Gamma} \cdot \mathbf{W}^{-1}, 
\]
\[
{\bf{G}}_{N \times (N+\delta+\gamma)}^{\rm noise} =\mathbf{W} \cdot \mathbf{K} \cdot \mathbf{P} \cdot \mathbf{V}^{\rm rx} \cdot \mathbf{R} . 
\]
Note that for $\nu \leq \gamma-\beta$, the number of \fcro{transmitted} data vectors affecting the reception (in the transform domain) of a single data vector is one. 
In this case, a set of $N$ independent parallel subcarriers is obtained, each having a channel gain of $H_{k}$, defined as the $N$-point DFT of the CIR $\bf{h}$. Thus a one-tap per subcarrier equalizer can be used to mitigate the phase and the amplitude distortion introduced by the channel. This means that ${\bf{A}}_m=0$, $m > 0$, and ${\bf{A}}_0$ is a diagonal matrix with elements $H_k$, $0\leq k \leq (N-1)$. 
For the other cases ($\nu>\gamma-\beta$), we have three different types of interference, as depicted in Fig. \ref{fig:Tipos-de-interferencias} \cite{Pol97, Mar19}:
\begin{itemize}
\item {Type-I intercarrier interference (ICI$_1$):} corresponding to the elements $\left[\mathbf{A}_0 \right]_{i,j}$, $i \neq j$.
\item {Type-II intercarrier interference (ICI$_2$):} corresponding to the elements $\left[\mathbf{A}_m\right]_{i,j}$, $i \neq j$, $m>0$. 
\item {Intersymbol interference (ISI):} corresponding to the diagonal elements  $\left[\mathbf{A}_m\right]_{i,i}$,  $m > 0$. 
\end{itemize}

We now derive theoretical expressions for the powers corresponding to the desired signal component in the received data vector, as well as to the ISI, ICI, and noise. These powers are used to compute the SINR. For this study, we assume that the components of the \fcro{transmitted} data vector $X_k$ and the noise vector $q_k$ are zero-mean wide-sense stationary uncorrelated processes, independent and identically distributed for all $k$, with variances $\sigma _X^2$ and $\sigma_n^2$, respectively.

\noindent \fcr{Let us rewrite \eqref{Eq_general_Yn} as
\begin{equation}\label{Eq_general_Yn2}
{\bf{Y}}[l]   = {\bf{A}}_0  \cdot {\bf{X}}[l] +\sum\limits_{m =  1}^M {{\bf{A}}_m  \cdot {\bf{X}}[l - m]} + {\bf{G}}^{\rm noise}  \cdot {\bf{q}}[l] .
\end{equation}
}
\noindent The desired signal component in the received data vector can be written as
\begin{equation}
\mathbf{Y}_{{\rm des}} [l] = \mathbf{A}_0^{{\rm{des}}} \cdot \mathbf{X}[l], 
\label{eq:Yr_des}
\end{equation}
where \diniz{  
${\mathbf{A}}_{0}^{\rm des}$ is a diagonal matrix with entries $\left[ \mathbf{A}_0^{{\rm{des}}} \right] _{i,i} = \left[ {\mathbf{A}}_0 \right]_{i,i}.$}
The desired signal power (before the transform-domain equalization) at subcarrier $k$ is obtained as the $(k,k)$-th element of the covariance matrix, i.e., $P_{\rm signal} \left( k \right) = \left[ {\bf{C}}^s \right]_{k,k}$, where
\fcr{\begin{eqnarray}\label{Ps_Yr_des}
 {\bf{C}}^s  &=& E\left\{ { \mathbf{Y}_{{\rm des}} [l]  \cdot \mathbf{Y}_{{\rm des}}^H } [l]\right\}  \nonumber \\ 
  &=& E\left\{ {{\bf{A}}_{0}^{\rm des} \cdot{\bf{X}}[l]\cdot{\bf{X}}^H [l] \cdot \left( {{\bf{A}}_{0}^{\rm des} } \right)^H } \right\} \nonumber \\ 
  &=& {\bf{A}}_{0}^{\rm des} \cdot E\left\{ {\bf{X}}[l]\cdot{\bf{X}}^H [l] \right\} \cdot \left( {{\bf{A}}_{0}^{\rm des} } \right)^H  \nonumber \\  
  &= &  \sigma _X^2 \cdot{\bf{A}}_{0}^{\rm des} \cdot \left( {{\bf{A}}_{0}^{\rm des} } \right)^H,  
\end{eqnarray}}
\noindent \fcro{where} $E\left\{ \cdot \right\}$ is the expected-value operator. The noise data vector is given by
\begin{equation}
\mathbf{Y}_{{\rm noise}} [l] =  \mathbf{G}^{\rm noise}\cdot\mathbf{q}[l],
\end{equation}
As a result, the noise power is given by $P_{\rm noise} \left( k \right)=\left[ {\bf{C}}^n \right]_{k,k}$, where
\fcr{\begin{eqnarray}\label{Pn_des}
{\bf{C}}^n & = &  {\bf{G}}^{\rm noise} \cdot E \left\{ \mathbf{q}[l] \cdot \mathbf{q}^H[l] \right\} \cdot \left( {{\bf{G}}^{\rm noise} } \right)^H \nonumber \\
& = &  \sigma _n^2 \cdot {\bf{G}}^{\rm noise} \cdot \left( {{\bf{G}}^{\rm noise} } \right)^H.  
\end{eqnarray}}
The interference component is given by
\begin{eqnarray}\label{eq:Yr_int}
 \mathbf{Y}_{{\rm int}} [l]  &=& \mathbf{Y} [l]-\mathbf{Y}_{{\rm des}} [l] - \mathbf{Y}_{{\rm noise}} [l]  \nonumber \\
&=& \mathbf{A}_{0}^{\rm ICI_1}\cdot \mathbf{X}[l] + \sum\limits_{m=1}^M {\mathbf{A}_{m} \cdot \mathbf{X}[l-m]},
\end{eqnarray}
where $\mathbf{A}_0^{\rm ICI_1} =\mathbf{A}_0-\mathbf{A}_0^{\rm des}$. Using the above, the ISI and ICI power is $P_{{\rm{ISI}},{\rm{ICI}}} \left( k \right)  = \left[ {\bf{C}}^i \right]_{k,k}$, where
\begin{align}\label{Pint_Yr}
 {\bf{C}}^i  & = E \left\{ {\mathbf{Y}_{{\rm{int}}} [l] \cdot \mathbf{Y}_{{\rm{int}}}^H [l] } \right\} \nonumber \\
& = E\left\{ {{\bf{A}}_{0}^{{\rm{ICI}}_{\rm{1}} } \cdot{\bf{X}}[l]\cdot{\bf{X}}^H [l]\cdot\left( {{\bf{A}}_{0}^{{\rm{ICI}}_{\rm{1}} } } \right)^H } \right\} \nonumber \\
& + \sum\limits_{m=1}^M E \left\{ {{\bf{A}}_{m} \cdot{\bf{X}}[l - m]\cdot{\bf{X}}^H [l - m]\cdot\left( {{\bf{A}}_{m} } \right)^H } \right\}{\rm{ }} \nonumber \\
 & =  \sigma _X^2 \cdot\left( {{\bf{A}}_{0}^{{\rm{ICI}}_{\rm{1}} } \cdot\left( {{\bf{A}}_{0}^{{\rm{ICI}}_{\rm{1}} } } \right)^H  + \sum\limits_{m=1}^M {{\bf{A}}_{m} \cdot\left( {{\bf{A}}_{m} } \right)^H } } \right).
\end{align}
\fcr{Finally}, the SINR for subcarrier $k$ is
\fcr{\begin{equation}\label{eqSINRDCT}
{\rm SINR}\left( k \right) = \frac{P_{\rm signal}\left( k \right)}
{P_{\rm ISI,ICI} \left( k \right) + P_{\rm noise} \left( k \right)} = \frac{\left[ {\bf{C}}^s \right]_{k,k}}
{\left[ {\bf{C}}^i \right]_{k,k} + \left[ {\bf{C}}^n \right]_{k,k}}.
\end{equation}}

\begin{figure}
\centering{
\subfigure[]{\includegraphics[width=0.45\textwidth]{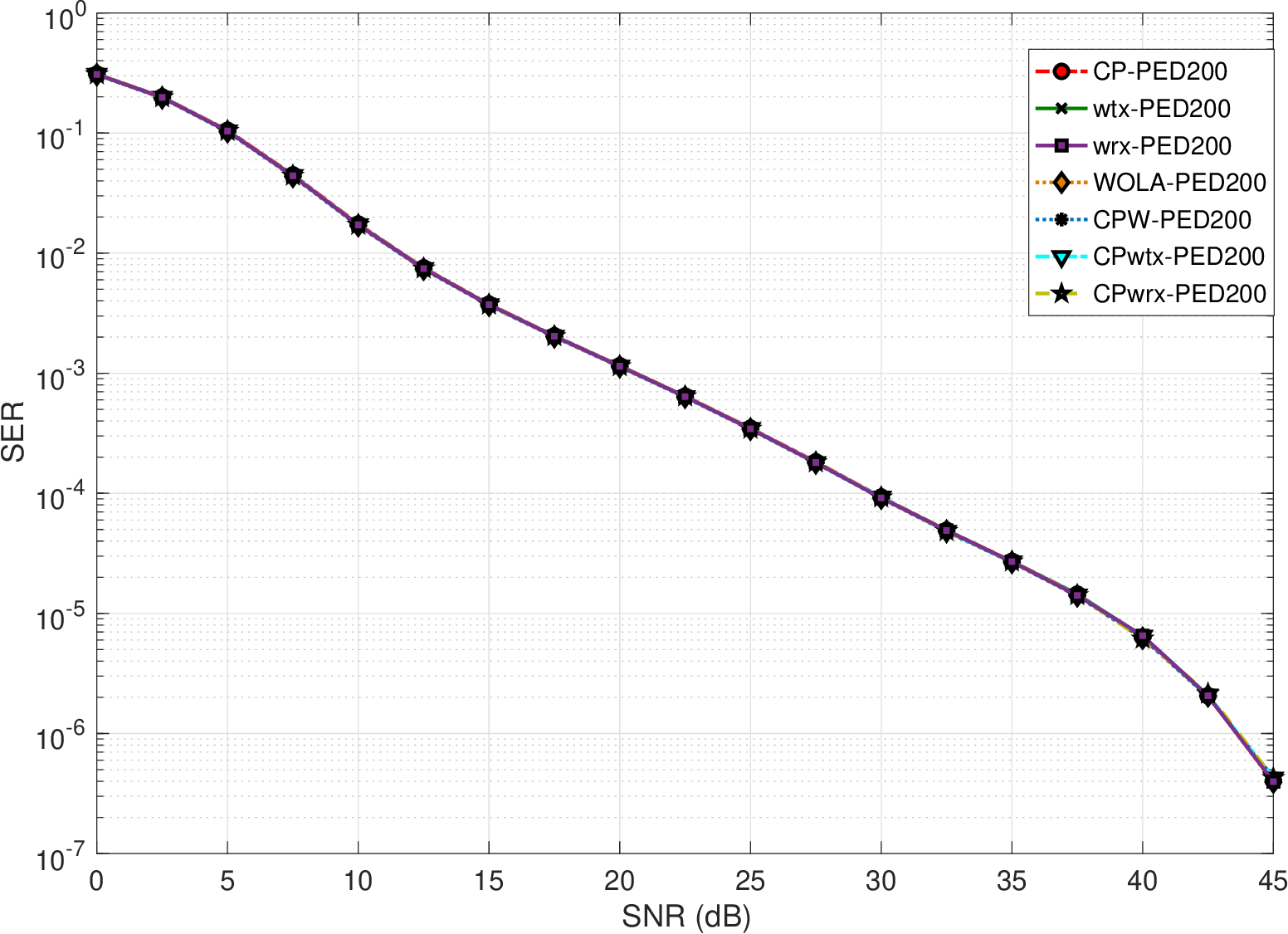}  \label{fig-SER_longi_PedA10}}  \hfill 
\subfigure[]{\includegraphics[width=0.45\textwidth]{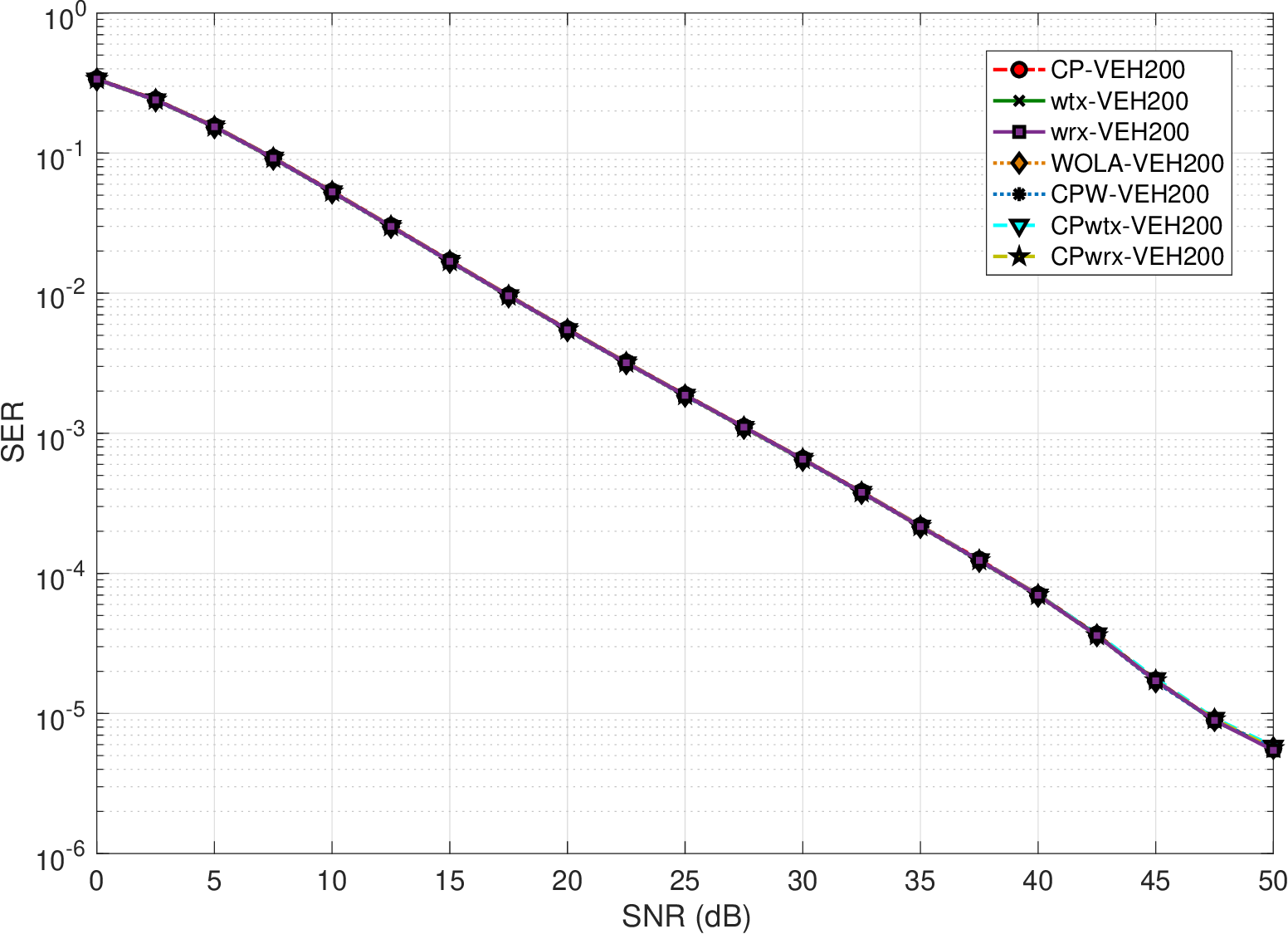} \hfill \label{fig-SER_longi_VehA200}}
} 
\caption{SER versus SNR for different OFDM systems. (a) PED200. (b) VEH200.}
\label{fig_SER_snr_values}
\end{figure}

\begin{table*}
\center{}
\caption{Parameters used in the experiments ($N=256$)} 
{\begin{tabular}{ | c | c | c |  c | c| c |c | c | }
  \hline
   \textbf{Parameter} & \textbf{CP-OFDM} & \textbf{wtx-OFDM} & \textbf{wrx-OFDM} &  \textbf{WOLA-OFDM} &  \textbf{CPW-OFDM} & \textbf{CPwtx-OFDM} & \textbf{CPwrx-OFDM}  \\
  \hline
  \hline
   $\beta$  & $0$ & $8$ & $0$ & $8$ & $8$  & $8$ & $0$  \\
   \hline
   $\delta$  & $0$ & $0$ &$10$ & $10$  & $10$  & $0$   & $10$   \\
   \hline
   $\rho$  & $0$ & $8$ & $5$ & $8$  & $13$   & $0$ & $0$ \\
   \hline
   $\gamma$  & $32$ & $32$ & $27$ & $22$ & $27$  & $24$ & $22$  \\
   \hline
    $\kappa$ & $0$  & $0$ & $0$  & $5$ & $0$  & $8$ & $5$  \\
   \hline
\end{tabular}}{\label{parameter_simulat}}
\end{table*}

\section{Simulations}\label{sec:experiments}
In order to demonstrate the applicability of the proposed \fcro{unified} formulation, this section compares the performance of the studied systems in terms of SER and achievable data rate. It is worth noting that OOB emissions will not be taken into account here. The set of parameters used in the simulations are summarized in Table \ref{parameter_simulat}. BPSK modulation is used as the primary mapping, the number of active subcarriers is $N = 256$, which is the DFT size, and the frequency spacing is $11.16071492$ kHz. Two sets of 250 wireless fading channels each, according to the ITU Pedestrian A and Vehicular A channels \cite{Itu97, 3GPP07}, are used as multipath channels. They have been generated with Matlab's \texttt{stdchan} using the channel models \texttt{itur3GPAx} and \texttt{itur3GVAx} with a carrier frequency $f_{\rm c}=2$ GHz and two different sets of parameters: (a) 4 km per hour as pedestrian velocity, $T_{\rm s}=200$~ns and length $L=\nu+1=11$; (b) 100 km per hour as mobile speed, $T_{\rm s}=200$~ns and  length $L=\nu+1=21$. These channels are referred to as PED200 and VEH200, respectively. The noise is \wam{modeled} as an additive white Gaussian noise. It is assumed that the channel remains unchanged within the same simulation and perfect channel estimation is performed at the receiver. Perfect time and frequency synchronization is also assumed. 

In Fig. \ref{fig_SER_snr_values}, the ${\rm SER}$  performance curves of the different OFDM systems and channels are depicted. As can be seen, the results for the different systems are practically indistinguishable for each set of channels. Thus, there is no clear advantage in terms of SER of any particular OFDM system over the other systems.  

\begin{figure}
\centering{
\subfigure[]{\includegraphics[width=0.45\textwidth]{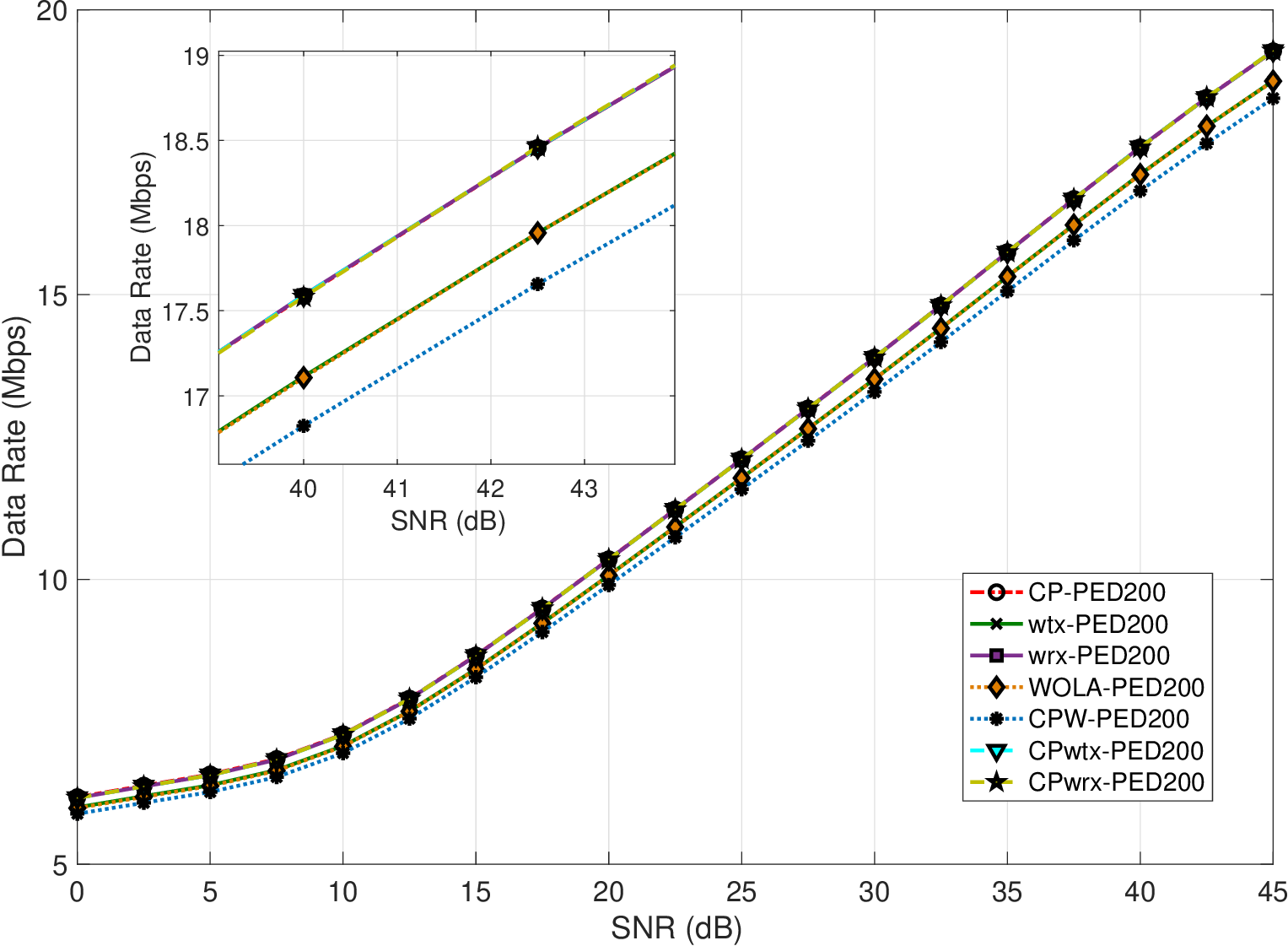}  \label{fig-Data_rate_PedA10}} \hfil 
\subfigure[]{\includegraphics[width=0.45\textwidth]{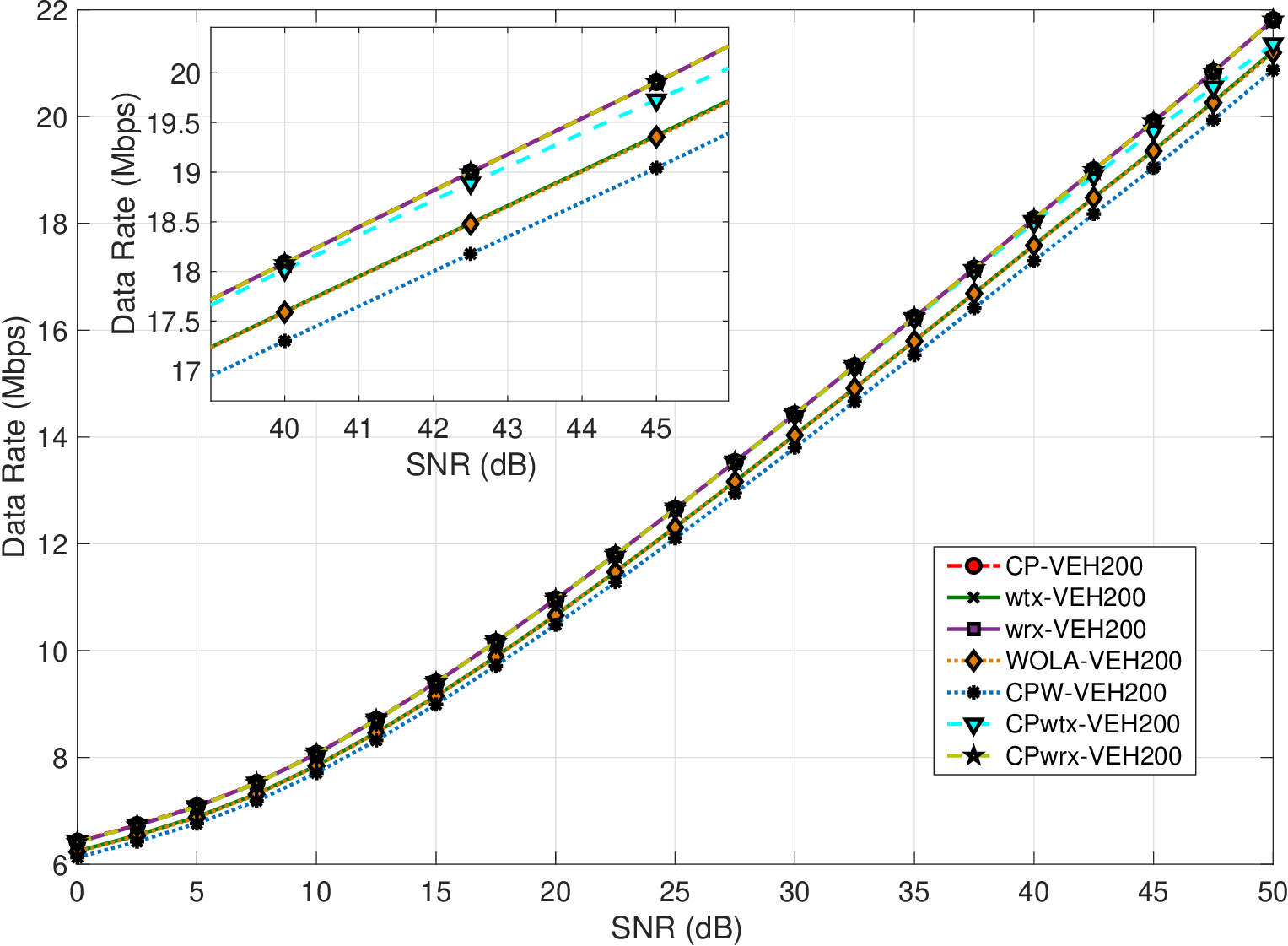}  \label{fig-Data_rate_VehA200}}
} 
\caption{Total achievable data rate versus SNR for different OFDM systems. (a) PED200. (b) VEH200.}
\label{fig_Data_rate_snr_values}
\end{figure}

\begin{figure}
\centering{
\subfigure[]{\includegraphics[width=0.45\textwidth]{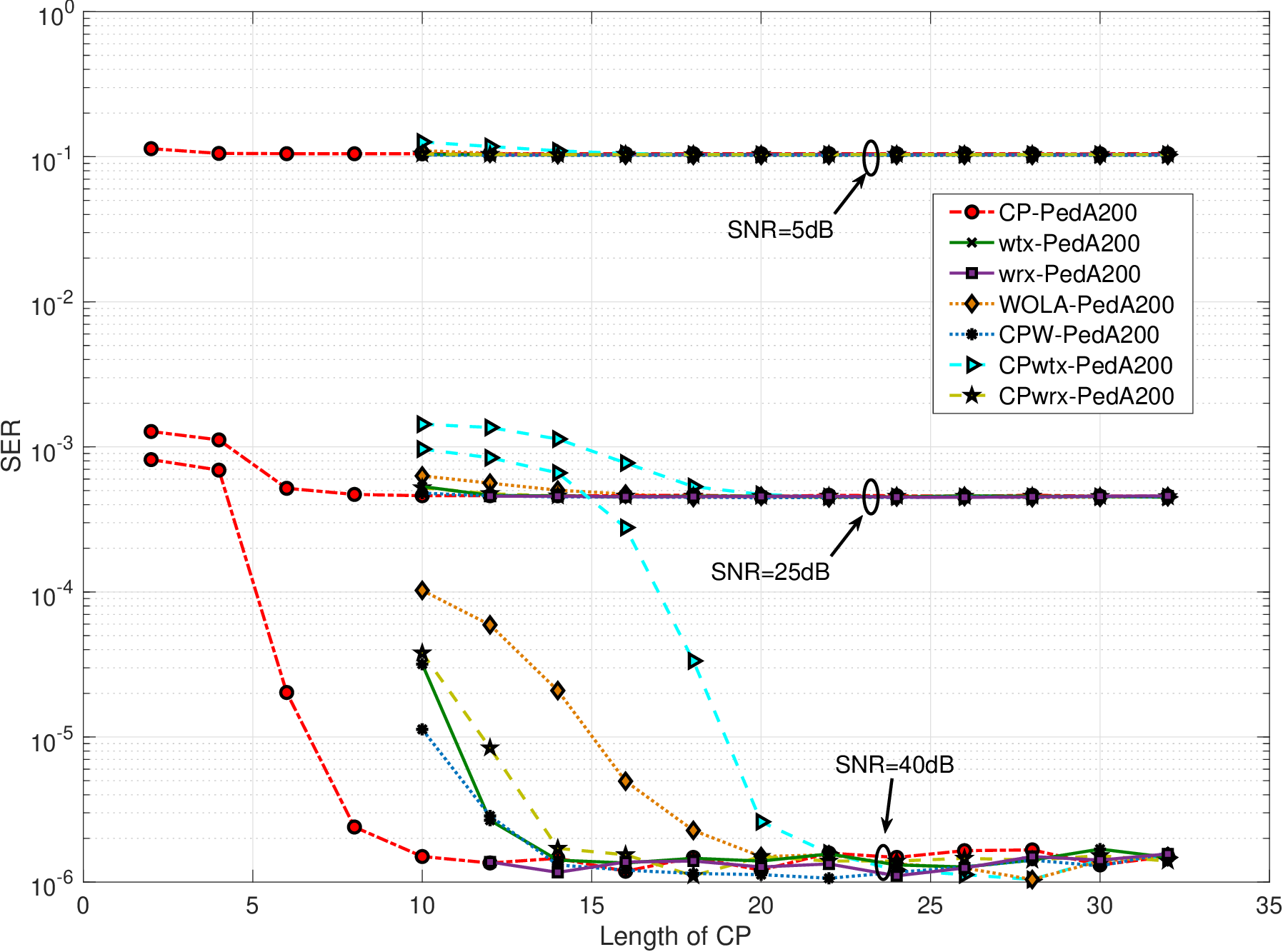}  \label{fig-SER_longi_CP_PedA10}} \hfil 
\subfigure[]{\includegraphics[width=0.45\textwidth]{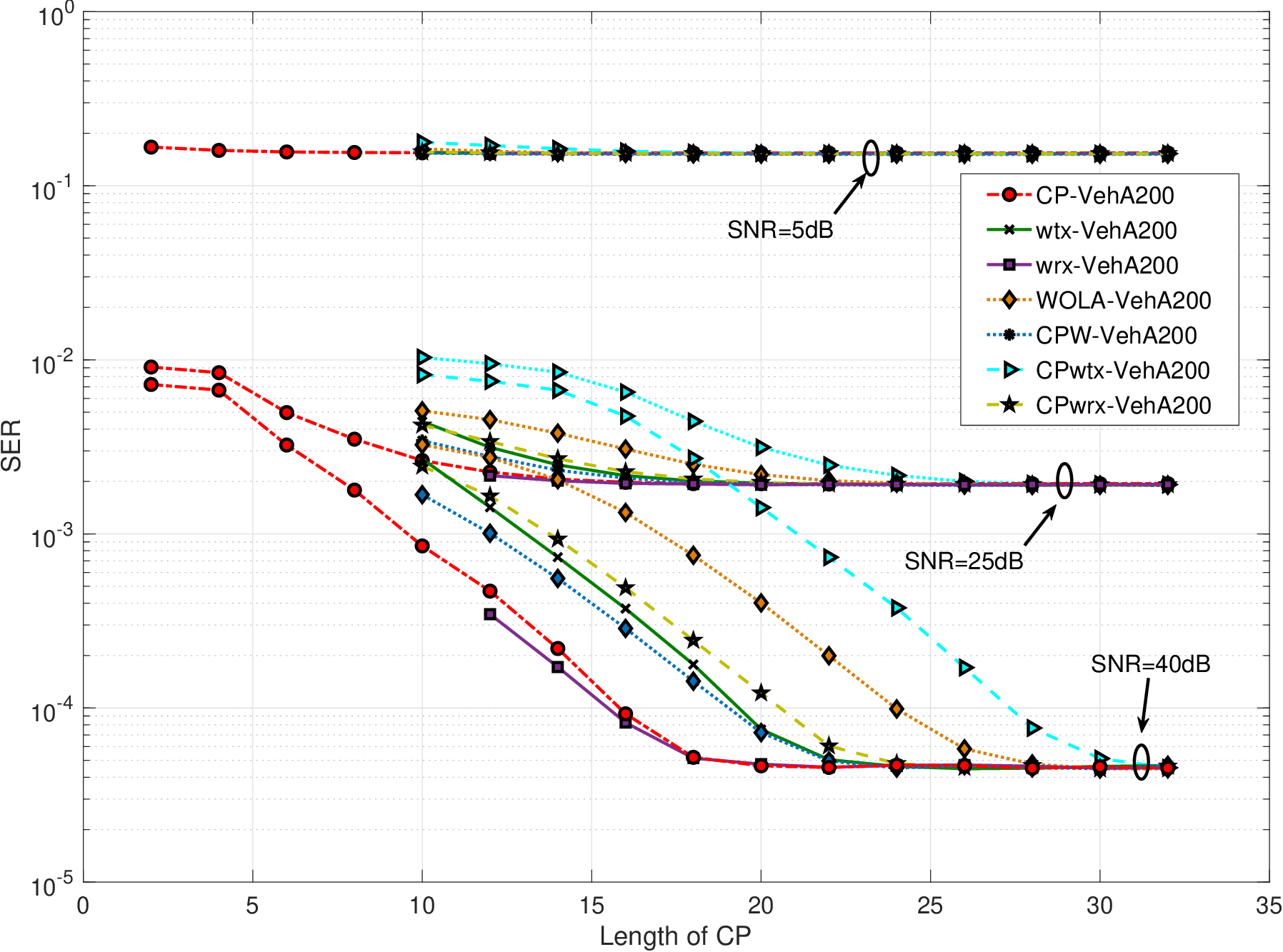}  \label{fig-SER_longi_CP_VehA200}}
} 
\caption{SER versus CP length for different OFDM systems. (a) PED200. (b) VEH200.}
\label{fig_SER_longi_CP_values}
\end{figure}

\begin{figure*}
\centering
\subfigure[]{\includegraphics[width=0.45\textwidth]{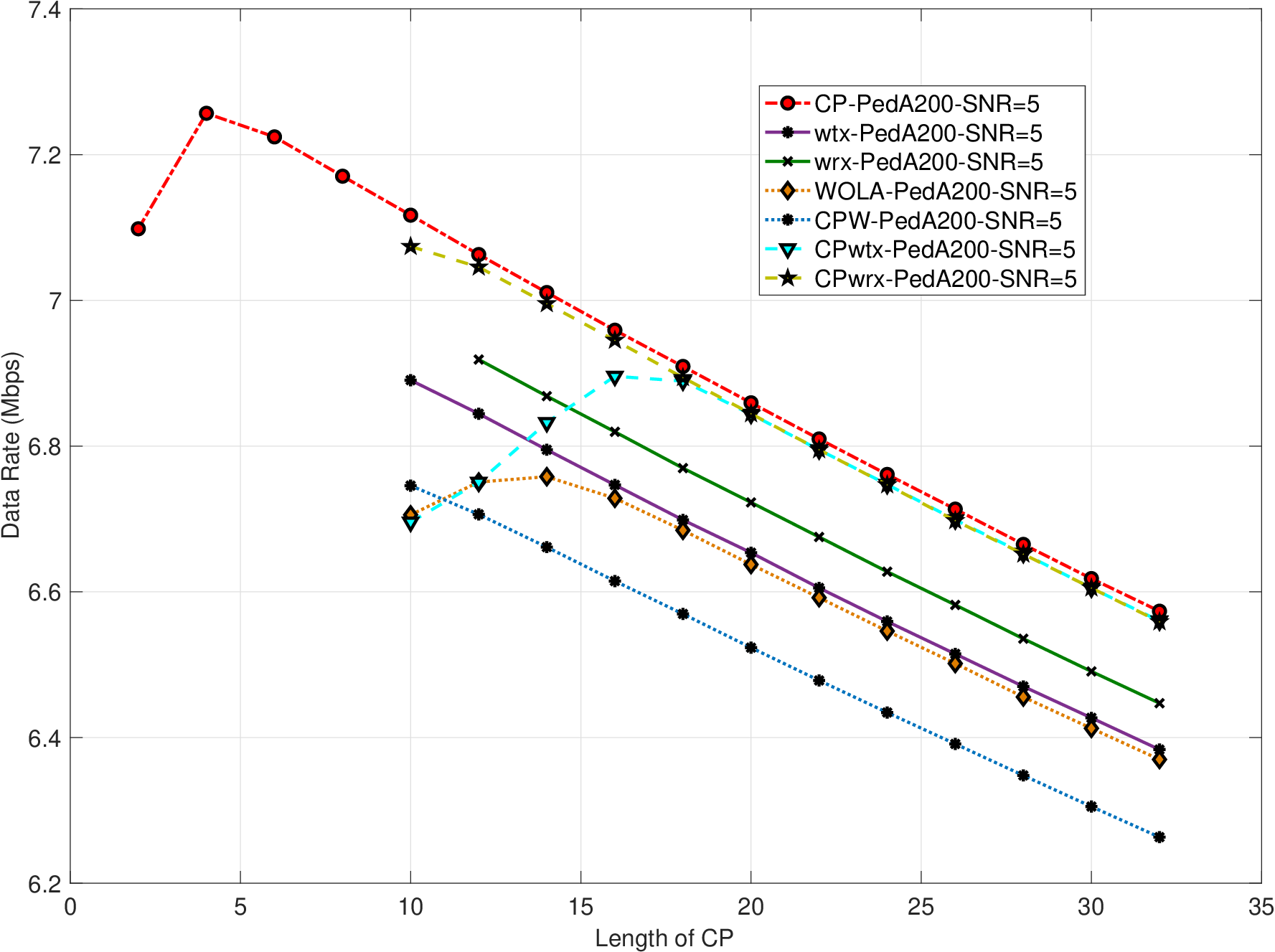}  \label{fig-longi_pedA200_SNR_5}}\hfill 
\vspace{0.5cm}
\subfigure[]{\includegraphics[width=0.45\textwidth]{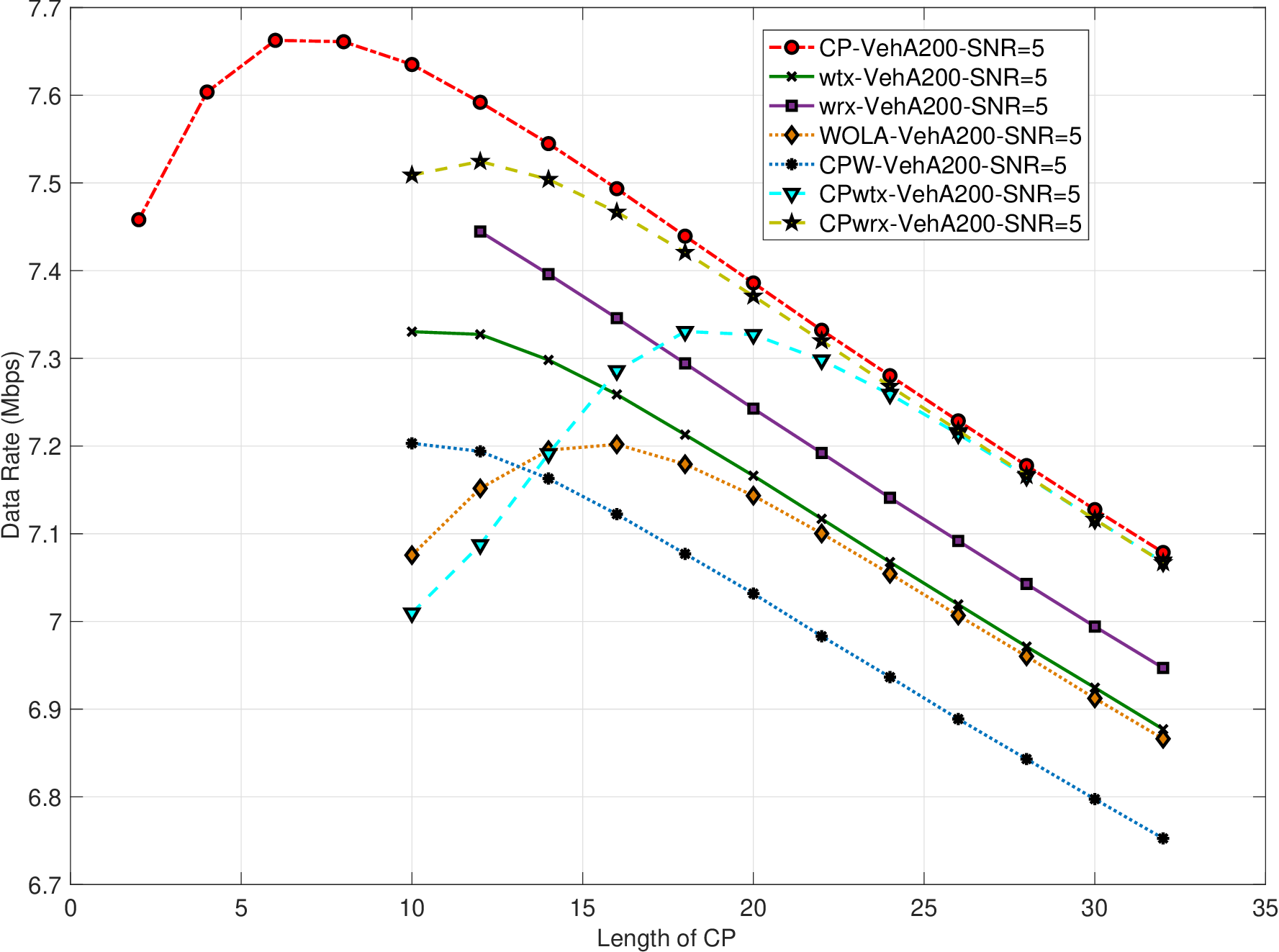}  \label{fig-longi_vehA200_SNR_5}}\hfill
\subfigure[]{\includegraphics[width=0.45\textwidth]{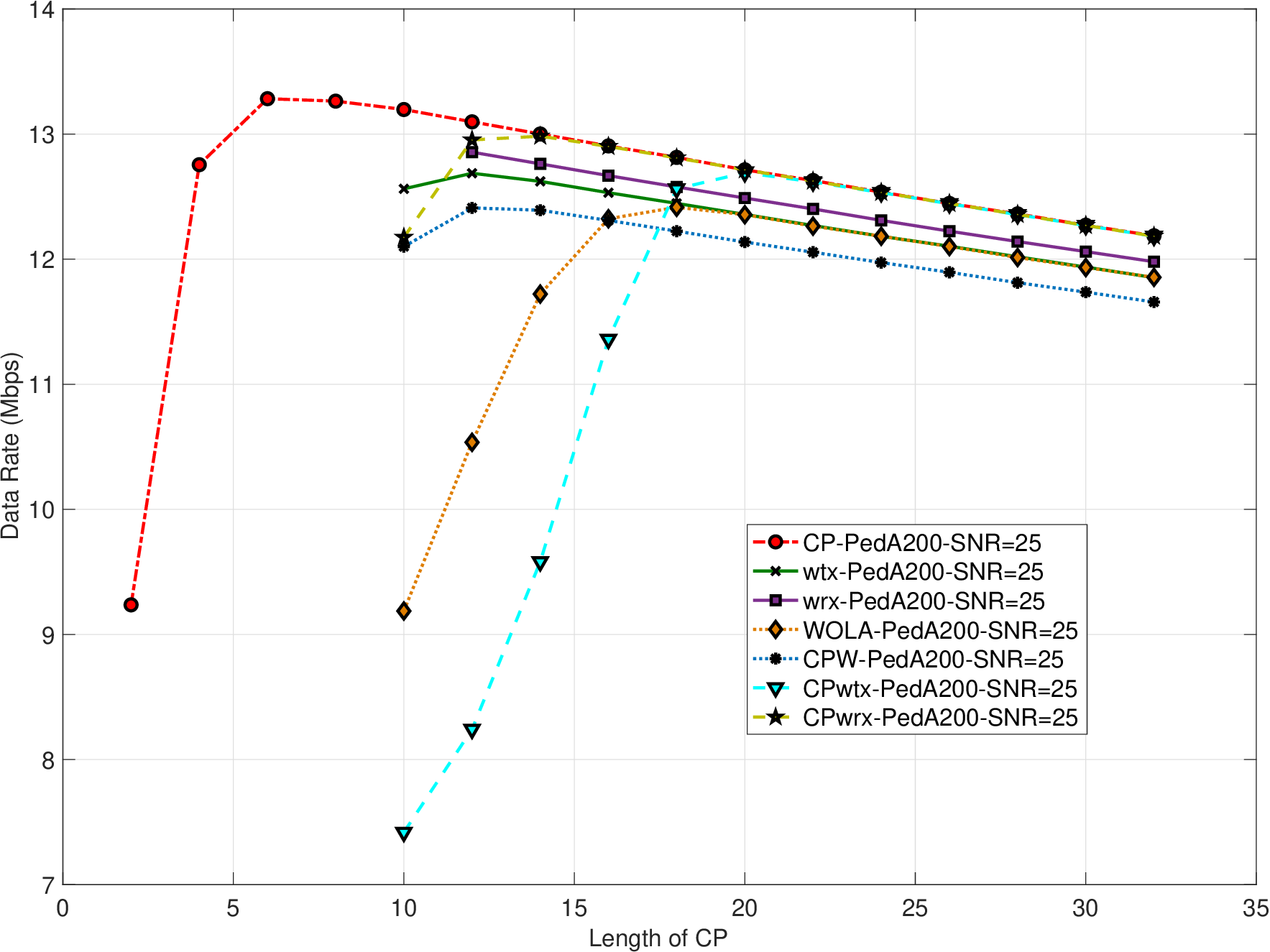}  \label{fig-longi_pedA200_SNR_25}}\hfill  
\subfigure[]{\includegraphics[width=0.45\textwidth]{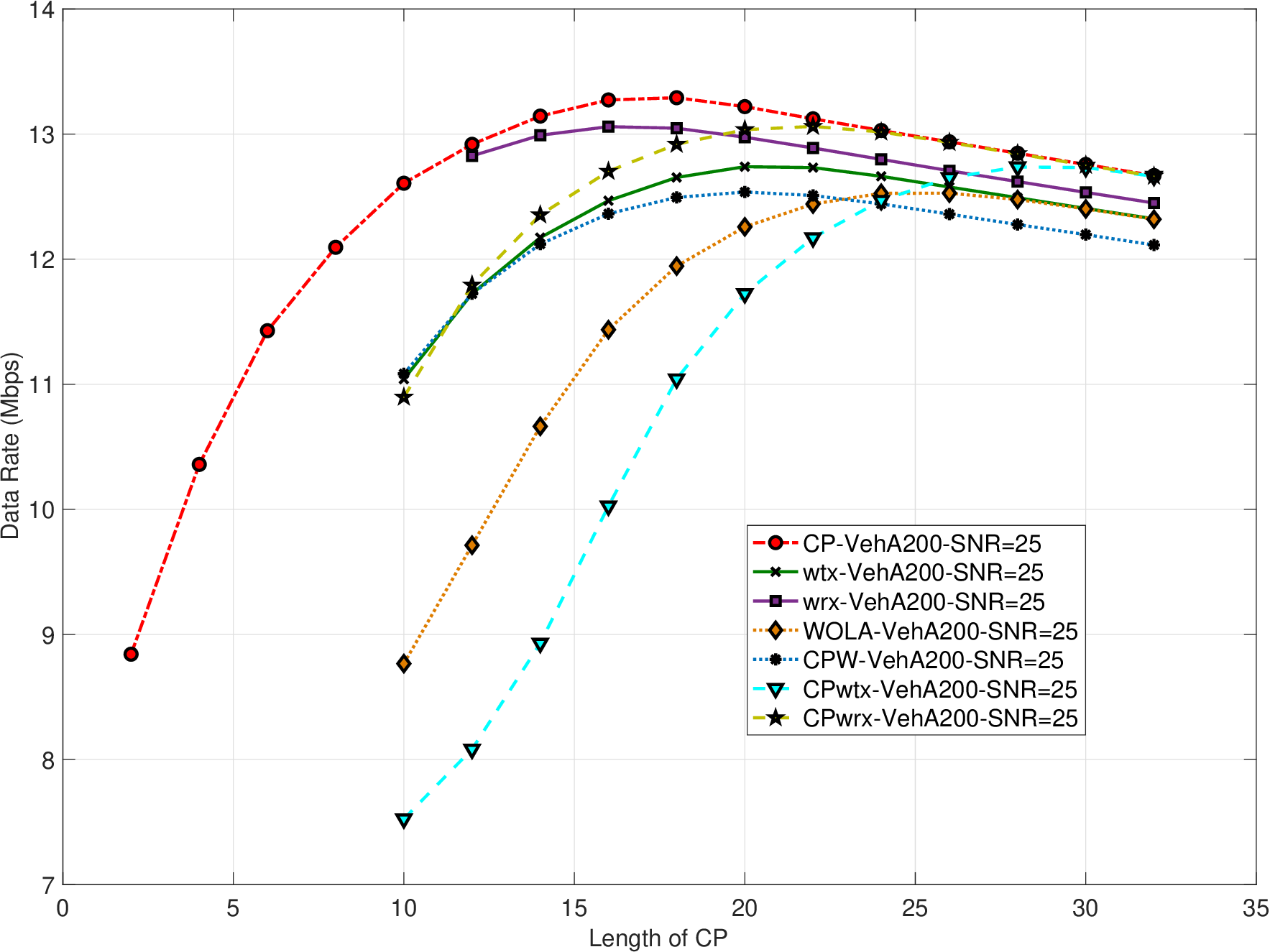}  \label{fig-longi_vehA200_SNR_25}}\hfill
\subfigure[]{\includegraphics[width=0.45\textwidth]{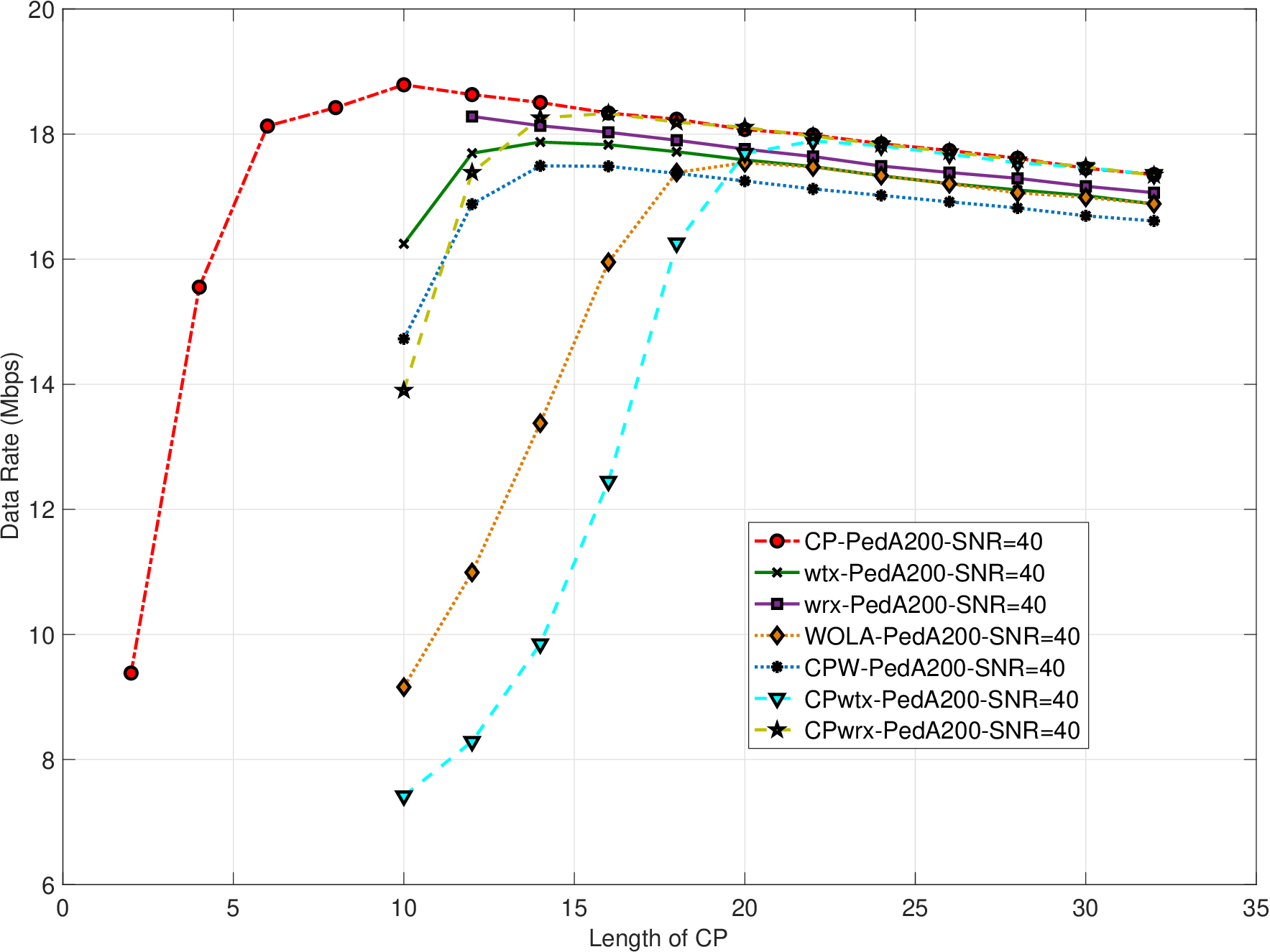}  \label{fig-longi_pedA200_SNR_45}}\hfill  
\subfigure[]{\includegraphics[width=0.45\textwidth]{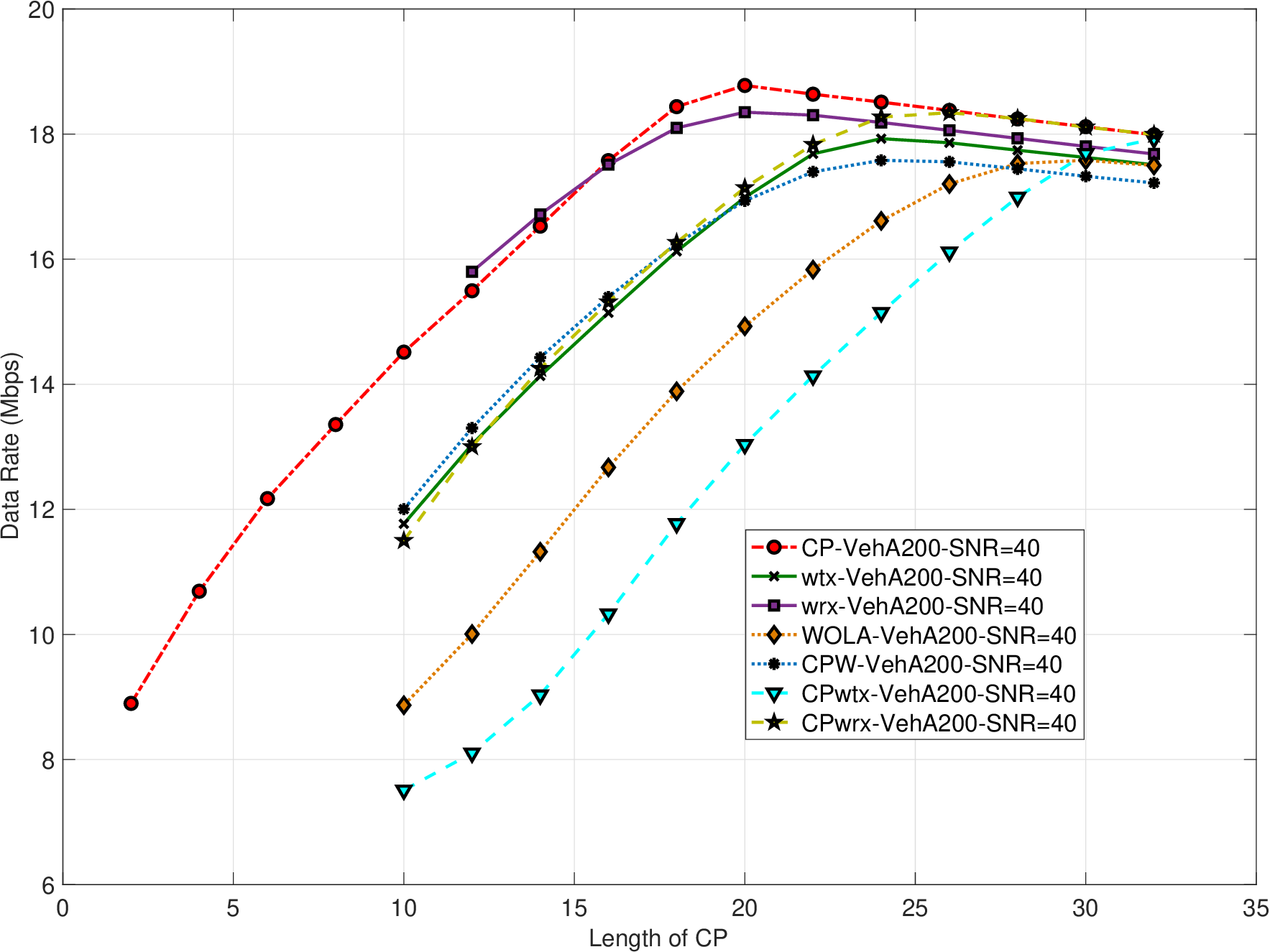}  \label{fig-longi_vehA200_SNR_45}}
\caption{Achievable data rate for different lengths for the CP and SNR values. }
\label{fig-data_rate_longi_snr_values}
\end{figure*}

Next, we investigate the data rate performance for a fixed CP length ($\mu=32$). As we use BPSK modulation, the data rate for subcarrier $k$  is given by \cite{Gar06}
\begin{equation}\label{eq_capacityDCT}
C \left( k \right)= \frac{1}{2} \log _2 \left( {\frac{{{\rm SINR}\left( k \right)}}{\gamma ^ {\ast} }} \right),
\end{equation}
where $\gamma ^ {\ast}$ is the modified ${\rm SINR}$ gap defined for a target SER as
\[
\gamma ^{\ast} = \left( {\frac{{Q^{ - 1} \left( {{{\rm SER} \mathord{\left/
 {\vphantom {{\rm SER} 2}} \right.
 \kern-\nulldelimiterspace} 2}} \right)}}{{\sqrt 2 \pi }}} \right)^2 .
\]
The total achievable data rate is thus
\begin{equation}\label{eq_throughputDCT}
R = f_{\rm s} \sum\limits_{k = 0}^{N - 1} {\frac{N}{{N_0}} \cdot C\left( k \right)} ,
\end{equation}
where $N_0=N+\mu+\rho$ and $f_{\rm s}={1 \mathord{\left/
 {\vphantom {1 {T_{\rm s}}}} \right.
 \kern-\nulldelimiterspace} {T_{\rm s}}}$. We employ the SER obtained in the previous simulations to compute the values of $\gamma ^{\ast}$ corresponding to each SNR. Fig. \ref{fig_Data_rate_snr_values} shows the resulting data rate as a function of the SNR. In this set of experiments, the OFDM systems that offer the best results are CP, wrx and CPwrx. The systems performing windowing in the \fcr{Rx} and 
including a prefix and suffix, such as WOLA and CPW, offer a
lower \fcro{data} rate due to the penalty of adding the two
types of redundant samples.

We now analyze the influence of the CP size on the resulting data rate. To this purpose, the SER as a function of the CP length is obtained for each OFDM system (see Fig. \ref{fig_SER_longi_CP_values}), assuming ${\rm SNR} = 5, 25$, and $40$~dB. These results are employed to calculate $\gamma^{\ast}$. Then, we obtain the achievable data rate, depicted in Fig. \ref{fig-data_rate_longi_snr_values} for the PED200 and VEH200 channels.  In all cases, CP-OFDM outperforms the other systems, except for ${\rm SNR} = 40$~dB, VEH200, and for smaller values of the CP, for which the \fcr{wrx-OFDM} shows a better performance. However, this improvement is not very significant in this case of insufficient redundant samples. The remaining OFDM systems have better performance whenever the windowing is in the \fcro{Rx}. For both small CP lengths and low SNR values, the systems that only have a CP outperform those that incorporate a CS.

\begin{figure}
\centering
\subfigure[]{\includegraphics[width=0.45\textwidth]{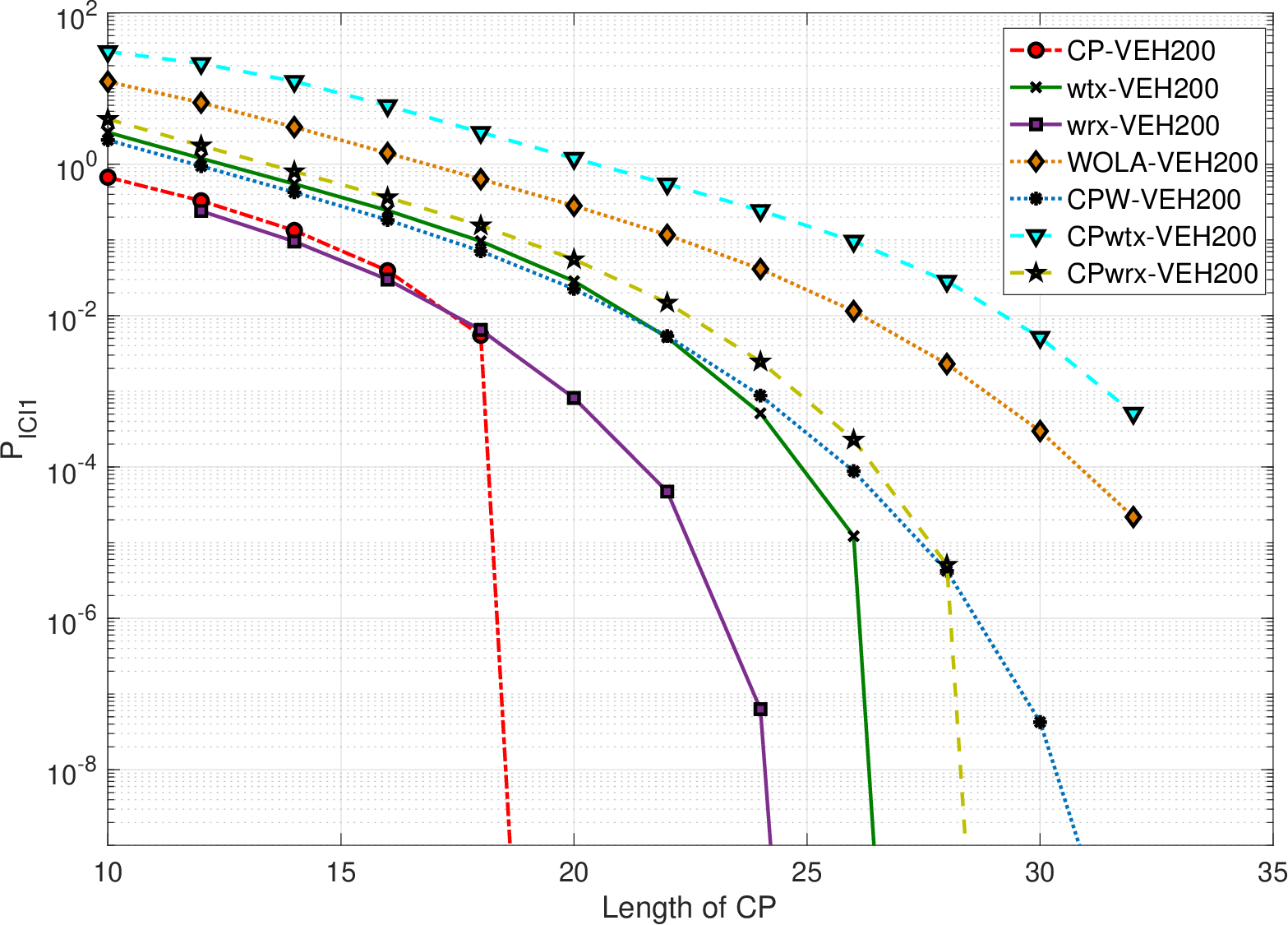}  \label{fig-PICI1_longi_vehA200_SNR_5}}\hfill
\subfigure[]{\includegraphics[width=0.45\textwidth]{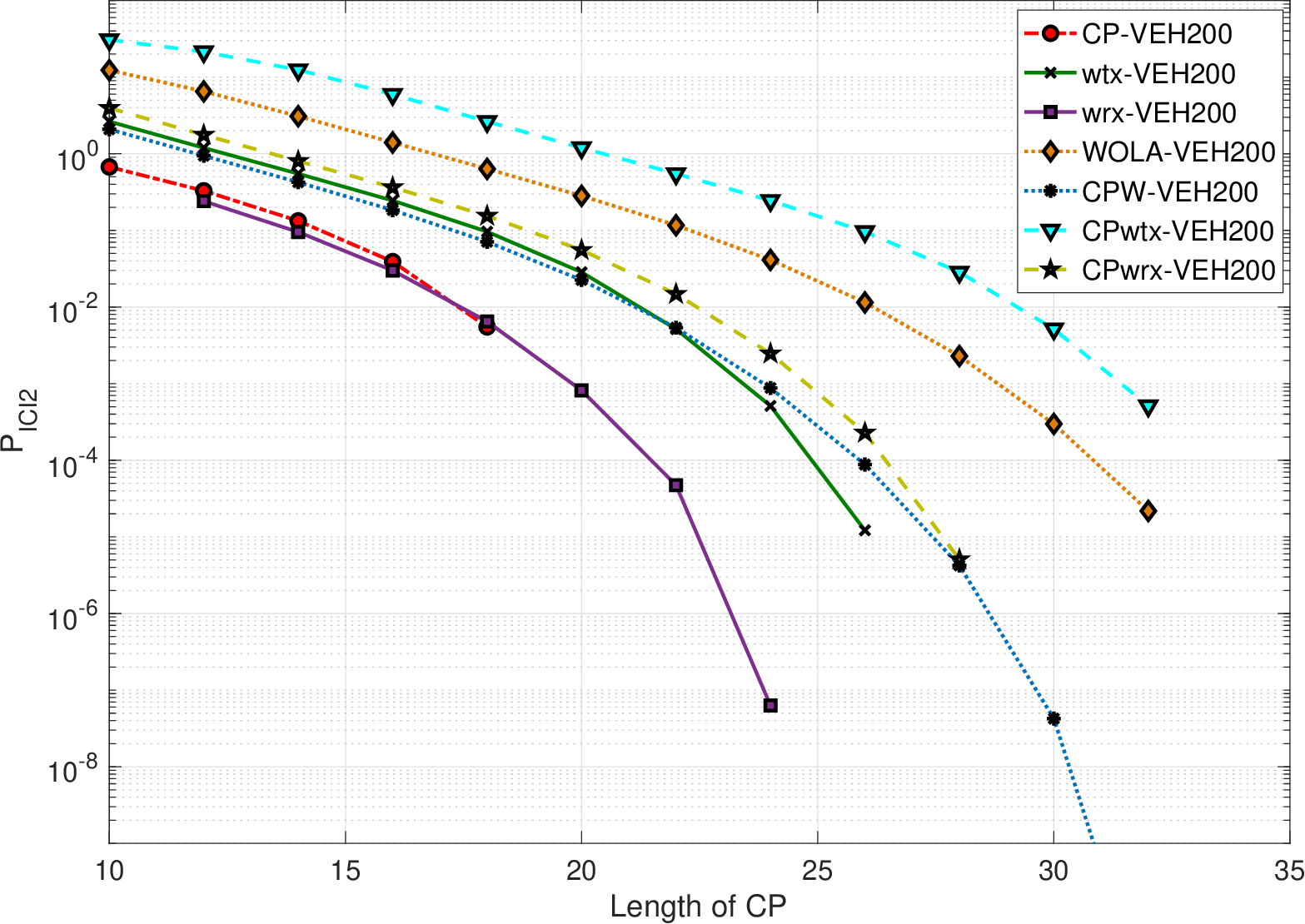}  \label{fig-PICI1_longi_vehA200_SNR_25}}\hfill
\subfigure[]{\includegraphics[width=0.45\textwidth]{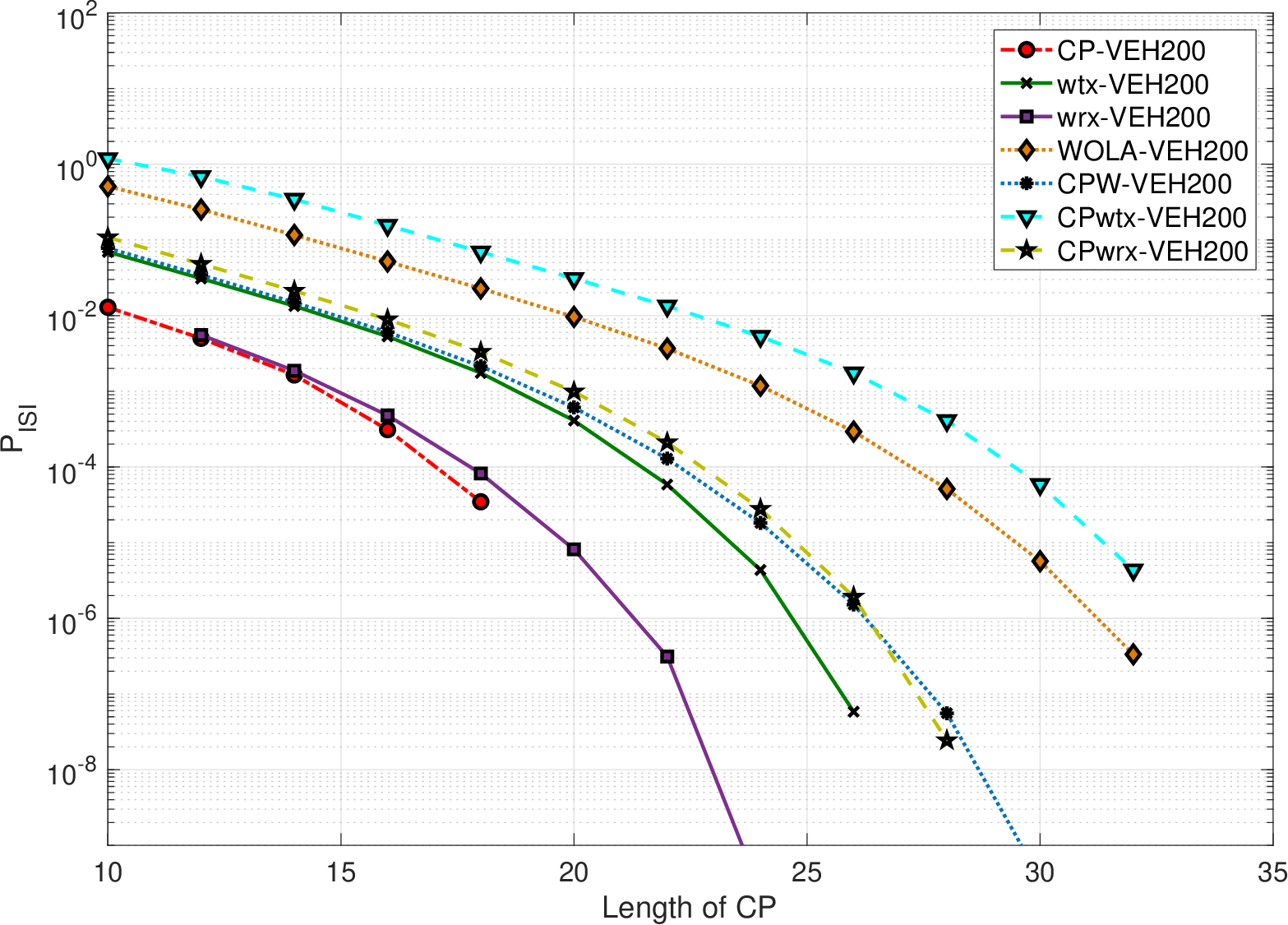}  \label{fig-PICI1_longi_vehA200_SNR_45}}
\caption{Total power for different CP lengths. (a) Power of type-I intercarrier interference. (b) Power of type-II intercarrier interference. (c) Power of intersymbol interference.}
\label{fig-Pvarios_longi_snr_values}
\end{figure}

Finally, the formulation presented here allows analysis in the transform domain of the three different interference powers that appear in each OFDM scheme. 
Fig. \ref{fig-Pvarios_longi_snr_values} shows the total power results ($P_{\rm ICI1}$, $P_{\rm ICI2}$, and $P_{\rm ISI}$) as a function of the CP length, obtained in the previous experiment for the VEH200 channel. The interference power is higher for systems whose windowing is performed in the \fcro{Tx} than those with windowing in the \fcro{Rx}. Note that \fcro{CPW-OFDM} has low levels of interference power, but the data rate results do not outperform the other systems. This is due to the overhead involved in the inclusion of both \fcro{a} CP and CS.
 
\section{Conclusion}\label{sec:conclusion}
In this paper, we \fcro{have} presented a unified formulation that describes conventional CP-OFDM and \fcro{six} other different w-OFDM systems. The \fcro{unified formulation} describes the whole transmitter, including the overlap-and-add windowing operation, and the operation of convolving the transmitted signal with the channel, as well as the complete receiver operation. Moreover, we \fcro{have} derived expressions for \fcro{the} intersymbol interference as well as two different kinds of intercarrier interference, along with their corresponding powers, besides the noise component. We \fcro{have} developed analytical expressions for the SINR so as to evaluate the effects of interference on the considered OFDM systems and to study the achievable data rate. Computer simulations \fcro{have been} carried out with practical scenarios. Comparing the obtained results, we \fcro{have} observed that in terms of SER, all OFDM systems behave similarly. \fcro{However, in terms of data rate}, the OFDM systems that only have a windowing in the receiver, or that only include a CP, outperform the \fcro{other systems}. It has also been noted that some systems (such as CPW-OFDM) have low interference power levels, but \fcro{their data rate performance is slightly lower compared to} other systems with more interference. The reason for this can be found in the penalty paid for including both the \fcro{CP and CS}.





\end{document}